\documentclass[12pt,aps,floatfix,pra,reprint,superscriptaddress,numerical,longbibliography]{revtex4-1}

\usepackage{dcolumn} 
\usepackage{bm}      
\usepackage{graphicx} 
\usepackage{xfrac}
\usepackage{extdash}
\usepackage{epsfig}
\usepackage{isomath}
\usepackage{textcomp}
\usepackage[english]{babel}
\usepackage{color,soul}
\usepackage{amsfonts}
\usepackage[columnwise]{lineno}
\usepackage{gensymb}
\usepackage{times}
\usepackage[normalem]{ulem}
\usepackage{xurl}
\usepackage[hidelinks]{hyperref}
\usepackage{amsmath}
\usepackage{capt-of}
\usepackage{hyperref}

\DeclareFontFamily{U}{euc}{}
\DeclareFontShape{U}{euc}{m}{n}{%
  <-6>eurm5 <6-8>eurm7 <8->eurm10}{} 
\DeclareSymbolFont{AMSc}{U}{euc}{m}{n} 
\DeclareMathSymbol{\umu}{\mathord}{AMSc}{"16} 

\newcommand{\ensuretext}[1]{\ensuremath{\text{#1}}}

\newcommand{\unit}[1]{\ensuretext{\textrm{\,}}\ensuremath{\mathrm{#1}}}

\renewcommand{\eqref}[1]{(\ref{#1})}

\begin{document}

\title{Reduction of SAXS Signal due to Doppler Broadening Induced Loss of Coherence} 

\author{Thomas Kluge}
\email{t.kluge@hzdr.de}
\affiliation{Helmholtz-Zentrum Dresden-Rossendorf, Bautzner Landstra\ss e 400, 01328 Dresden, Germany}

\author{Uwe Hernandez Acosta}
\author{Klaus Steiniger}
\affiliation{Center for Advanced Systems Understanding, Untermarkt 20, 02826 Görlitz, Germany}
\affiliation{Helmholtz-Zentrum Dresden-Rossendorf, Bautzner Landstra\ss e 400, 01328 Dresden, Germany}
\author{Ulrich Schramm}
\author{Thomas E. Cowan}
\affiliation{Helmholtz-Zentrum Dresden-Rossendorf, Bautzner Landstra\ss e 400, 01328 Dresden, Germany}
\affiliation{Technical University Dresden, Dresden, Germany}

\date{\today}

\begin{abstract}
We present an analytical and numerical study of how Doppler-induced spectral broadening in laser-heated plasmas degrades the coherence of small-angle X-ray scattering (SAXS) signals, and show that the resulting loss of temporal coherence reduces the SAXS intensity. 
Applying this formalism to two benchmark geometries— single density steps (wires) and periodic gratings—we obtain analytic estimates. For gratings, finite coherence simultaneously lowers Bragg-peak heights and broadens their widths, whereas for isolated steps only the overall scaling with $q$ affected. We map the parameter space relevant to current SASE and self-seeded XFELs, revealing that Doppler effects remain managable for the trieval of geometry parameters (less than few $10\,\%$ error) for SASE bandwidths but become the dominant error source in seeded configurations or above-keV temperatures. Practical consequences for density-gradient retrieval and interface-sharpness measurements are quantified. The results supply clear criteria for when Doppler broadening must be included in SAXS data analysis and offer a route to infer electron temperature directly from coherence-loss signatures.
\end{abstract}
\maketitle 

\section*{Introduction}
Small angle X-ray Scattering (SAXS) is increasingly becoming an important diagnostic tool for laser-generated high energy density plasmas. 
The reason lies in its high spatial resolution that can be used to trace the sharpness of phase transitions\cite{Kluge2017}, shocks\cite{Kluge2023} or ablation layers\cite{Kluge2018}, or to observe correlations at interfaces from roughness or instabilities\cite{Randolph2022,Randolph2024} or correlations between laser generated electron structures such as filaments or plasmons\cite{Ordyna2024}, at a very short time scale limited only by the duration and wavelength of the X-ray pulse and signal-to-noise level. 
Several examples of recent experiments where especially in high density plasmas the features of interest are beyond the resolution of current imaging methods demonstrate the need for this diagnostic, see e.g. \cite{Raj2020, LasoGarcia2024, schoenwaelder2024ultrafast}. 
SAXS as implemented at current XFELs has a sensitivity from a few to hundreds of nanometers \cite{Ordyna2024,Kluge2023,Gorkhover2016} and thus complements X-ray diffraction at atomic resolution and X-ray imaging at a few hundred nanometer resolution. However, precision experiments concerned with the sharpness of interfaces in the few nanometer range are often challenging as many effects can influence the results. 
For example, even extremely thin surface contaminants can lead to asymmetries in the scattering pattern ($I(q) \ne I(-q)$) and can therefore have significant effects for the density reconstruction, e.g. the sharpness of grating ridges, and can even be used to estimate the local plasma temperature \cite{Kluge2016,Gaus2021}. 

Another effect so far not extensively treated in literature is the possible influence of finite X-ray energy bandwidth and its change in laser-heated plasmas due to Doppler broadening. 
The question then naturally arises whether experimental deviations from the calculated scattering patterns neglecting these Doppler effects might be due to such temperature effects. \begin{figure}
    \centering
    \includegraphics[width=1\linewidth]{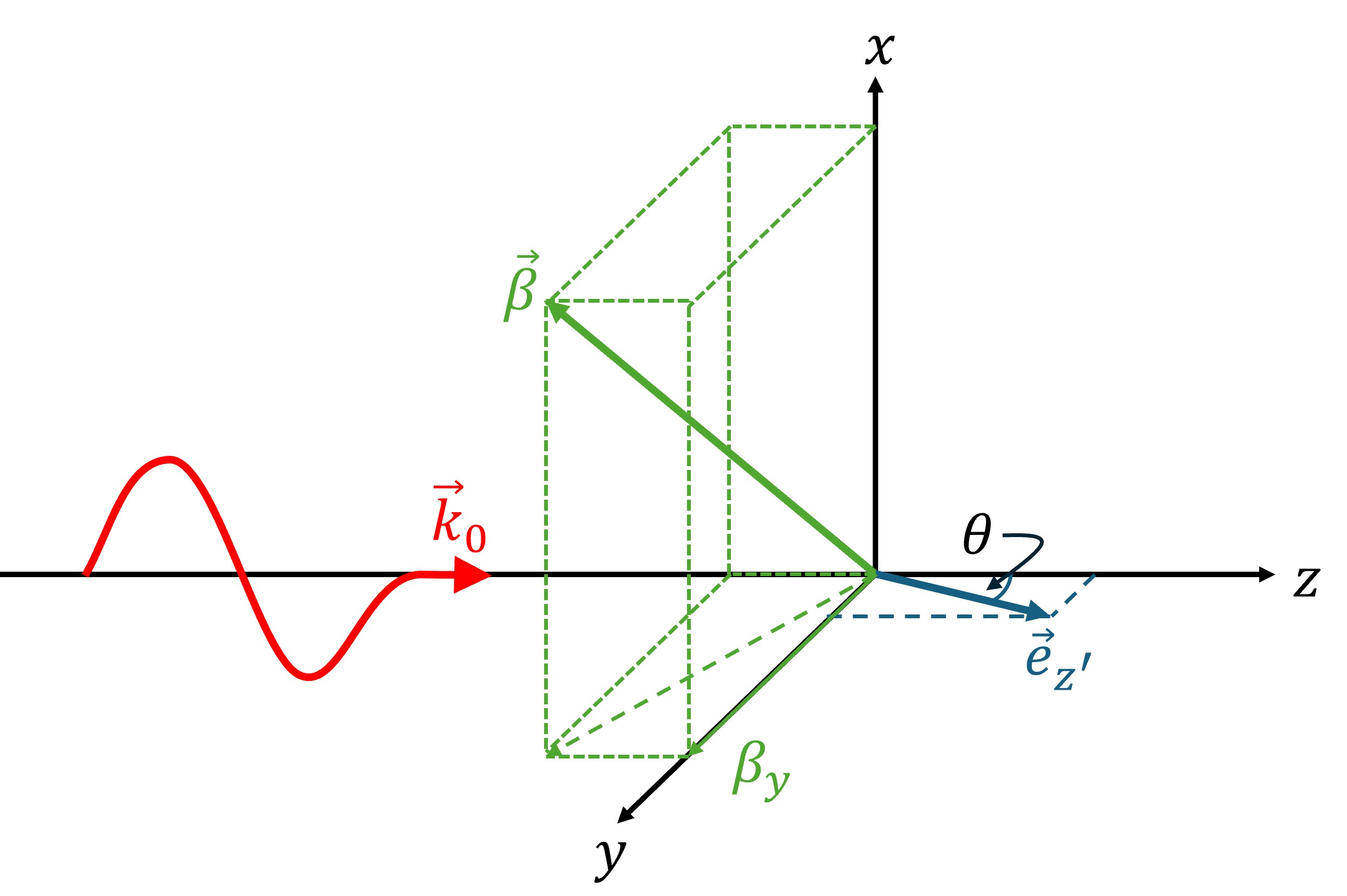}
    \caption{Definition of directions.}
    \label{fig:setup}
\end{figure}

In this paper we therefore want to investigate the relevance of Doppler related finite temperature effects on X-ray small angle scattering. 
We start by a quick recap of the Doppler shift in warm plasmas in Sec.~\ref{sec:Doppler} and its manifestation as effective spectral broadening when discussing coherence effects in Sec.~\ref{sec:Coherence}. 
The main focus of this paper is to then derive analytic estimates for the specific exemplary cases of gratings and step targets in Sec.~\ref{sec:Examples}, to give some quantitative approximations for practically relevant cases, and to analyze the importance of these considerations for the usual fitting procedure (Appendix C). 

\section{Doppler Shift in Warm Plasmas}
\label{sec:Doppler}
We consider a plasma with temperature $T \ll m_e c^2$. 
This plasma is irradiated by photons with frequency $\omega_0$ (i.e. energy $E_0$) and wave vector $\vec{k}_0=k_0\hat{\vec{e}}_z$ and we are interested in their frequency shift after scattering on the electrons. 

Let us first transform into the rest frame of an electron moving with velocity $\vec{\beta}$. 
Then the X-ray frequency seen by the electron is 
\begin{equation}
\omega_0' = \gamma\,\omega_0\,(1-\vec{\beta}\cdot\hat{\vec{e}}_z).
\label{eqn:e-restframe}
\end{equation}
The scattered photon energy in the electron rest frame is given by
\begin{equation}
\omega_{\text{out}}' = \frac{\omega_0'} {1+\frac{\hbar\omega_0'}{m_e c^2}\,(1-\cos\theta')}.
\end{equation}
We now assume without loss of generality that the observation direction $\hat{\vec{e}}_{obs}$ lies in the $x$--$z$ plane forming an angle $\theta$ with the $z$-axis. 
The scattered photon frequency in the lab frame then becomes with the observation direction $\hat{\vec{e}}_{obs^\prime} = \hat{\vec{e}}_{z}\,\cos{\theta^{\prime}}+ \hat{\vec{e}}_{x}\,\sin{\theta^\prime}$ (see Fig.~\ref{fig:setup} for the definition of directions)
\begin{equation}
\omega_{\text{out}} = \gamma\,\omega_{\text{out}}' \left(1+\vec{\beta}\cdot\hat{\vec{e}}_{obs^\prime}\right).
\end{equation}
Since the electron recoil can be neglected for $\hbar\omega_0 \ll m_e c^2$,
it is $\omega_{out}^\prime \approx \omega_0^\prime$ and hence
and with Eqn.~\eqref{eqn:e-restframe}
\begin{equation}
\omega_{\mathrm{out}}=\gamma^2\,\omega_0\,(1-\vec{\beta}\cdot\hat{\vec{e}}_z)\,(1+\vec{\beta}\cdot\hat{\vec{e}}_{obs^\prime})
\label{eqn:omega_0}
\end{equation}
With the aberration equation \cite{aberrationsgleichung}
\begin{equation}
    \cos{\theta^\prime}=\frac{\cos{\theta}-\beta}{1-\beta\,\cos{\theta}}
\end{equation}
this can be reritten as
\begin{equation}
    \omega_{\mathrm{out}}=\omega_0\frac{1-\vec{\beta}\cdot\hat{\vec{e}}_z}{1-\vec{\beta}\cdot\hat{\vec{e}}_{obs}}
\end{equation}
For small $\beta \ll 1$ this is 
\begin{equation}
    \omega_{\mathrm{out}}\cong\omega_0\,(1-\vec{\beta}\cdot\hat{\vec{e}}_z)(1+\vec{\beta}\cdot\hat{\vec{e}}_{obs})
\end{equation}
which is simply Eqn.~\eqref{eqn:omega_0} with $\gamma=1$ and $\hat{\vec{e}}_{obs^\prime}=\hat{\vec{e}}_{obs}$, i.e. $\theta^\prime=\theta$. 
In fact, even for few $10\,\mathrm{keV}$ photons and few $10\,\mathrm{keV}$ plasma temperature the error compared to the exact treatment is generally of the order of only few $\mathrm{meV}$ or less. 
In small angle approximation $\theta\ll1$ we can further simplify this with
\begin{equation}
\hat{\vec{e}}_{z^\prime} \approx \hat{\vec{e}}_z + \theta\,\hat{\vec{e}}_x,
\end{equation}
such that
\begin{equation}
\hat{\vec{e}}_{z^\prime}-\hat{\vec{e}}_z\approx \theta\,\hat{\vec{e}}_x\,,\,
\left(\vec{\beta}\cdot\hat{\vec{e}}_{z^\prime}\right)\left(\vec\beta\cdot\hat{\vec{e}}_{z}\right) < \beta^2. 
\end{equation}
Keeping only terms up to first order in   $\beta$ 
we finally obtain
\begin{equation}
\omega_{\text{out}}-\omega_0 = \omega_0\,\beta_y\,\theta.
\end{equation}
Thus, only the projection of the electron velocity onto the transverse plane in observation direction contributes. 

\begin{figure}
    \centering
    \includegraphics[width=1\linewidth]{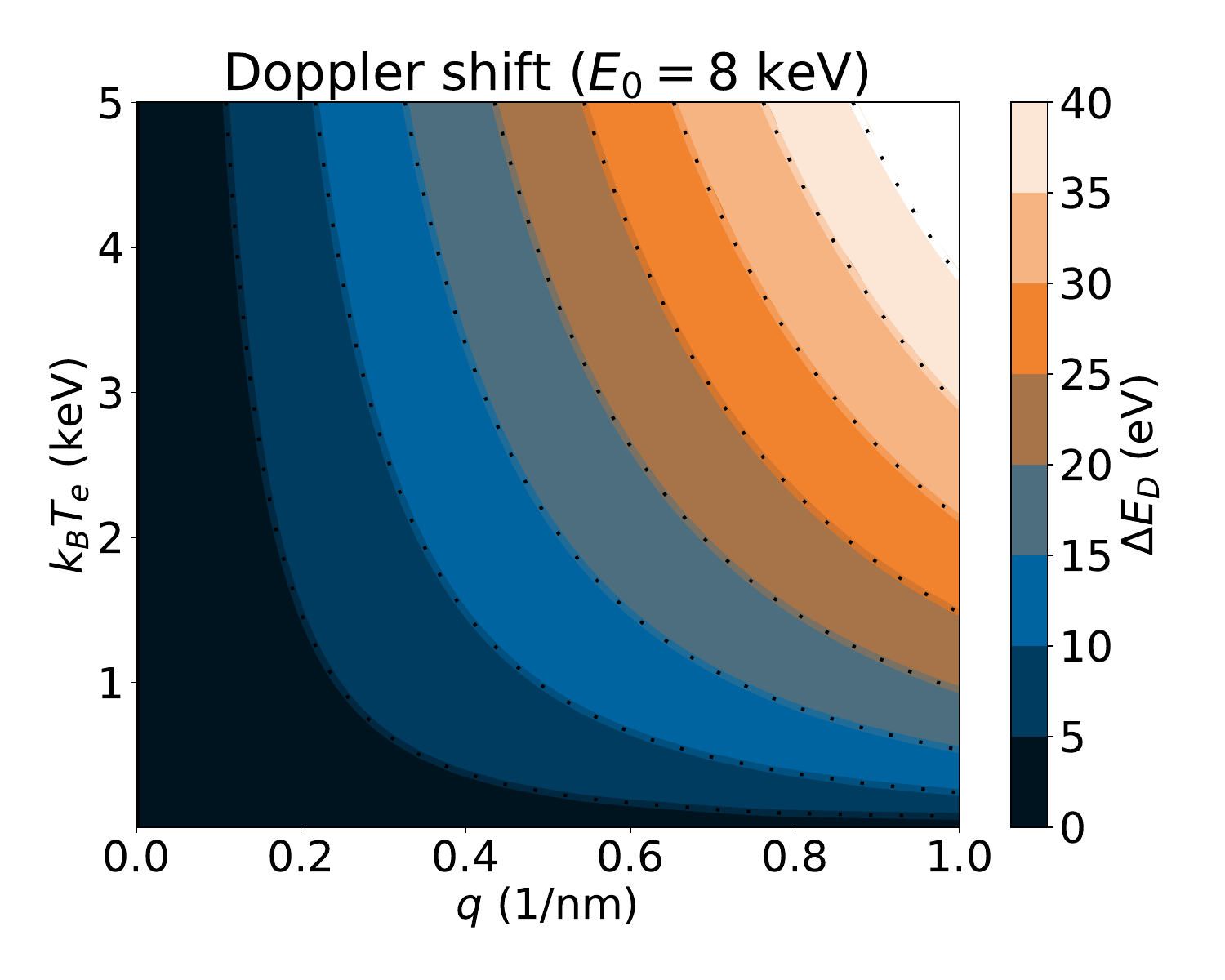}
    \caption{Doppler broadening FWHM from Eqn.~\eqref{eqn:doppler_shift} (solid lines with black dots) and the exact numerical solution including finite $\gamma$, electron recoil and exact Lorentz transformation of the observation angle $\theta$ into the electron rest frame, as well as Maxwell-Juettner diustributed electrons (filled color plot).}
    \label{fig:doppler_shift}
\end{figure}

We now want to compute the average Doppler shift 
\begin{equation}
\sigma_{\omega,D} \equiv \sqrt{\langle\left(\omega_{\text{out}}-\omega_0\right)^2 \rangle}. 
\end{equation}
For isotropically thermally distributed electrons and $\gamma\cong 1$ we can assume a Maxwell-Boltzmanm distribution
and therefore $\langle\beta_y^2\rangle = \frac{1}{3}\beta_{th}^2$ with  thermal velocity component in each direction
$\beta_{th}=\sqrt{3k_B T/m_e c^2}$. 
Then one obtains for the frequency variance
\begin{equation}
\sigma_{\omega,D}^2 \equiv \left\langle\left(\omega_{\text{out}}-\omega_0\right)^2\right\rangle = \omega_0^2 \, \frac{k_BT}{m_ec^2}\,\theta^2.
\end{equation}
Thus, the final expression for the Doppler induced spread of the cenral energies $\sigma_{E,D}=\hbar\sigma_{\omega,D}$ for low energy photons scattered on thermal electrons reads
\begin{equation}
\begin{aligned}
\sigma_{E,D} &= E_0\,\sqrt{\frac{k_B T}{m_e c^2}}\,\theta \\
         &\approx 0.28\,\mathrm{eV}\,\sqrt{T\left[\mathrm{eV}\right]}\cdot q\,\left[\frac{1}{\mathrm{nm}}\right]
         \label{eqn:doppler_shift}
\end{aligned}
\end{equation}
with the scattering vector $\vec{q} \cong \vec{e}_\theta \omega_0\sin{\theta}/c$ in small angle approximation. 
In other words for a thermal plasma the energy of an individual electron is shifted on average by $\pm\sigma_{E,D}$. 
In Fig.~\ref{fig:doppler_shift} we plot 
$E_D$ as a function of $T$ and $q$.\\
The same reasoning can be applied for the photon wave packet's spectral width $\sigma_{\omega,0}$, $\Delta\sigma_\omega=\sigma_{\omega,0}\,\sqrt{\frac{k_B T}{m_e c^2}}\,\theta$ which is much smaller than the shift of the central energy and will be neglected later. 

\section{Reduction of Scattering Signal Strength for Doppler Shifted Photon Beams}
\label{sec:Coherence}
In the following we assume an XFEL beam created by a stationary random process. 
SASE XFEL beams typically have a rather large bandwidth, as it consists of a train of several slices at different central energy, which we can treat as coherent within the respective slices. 
We now only treat a single coherent slice and assume that the individual slices add up incoherently. 
SASE XFEL beams have a relatively short coherence time, e.g. for $8\unit{keV}$ $\tau_{c,0}<0.4\unit{fs}$. 
As the full SASE bunch consists of multiple of these coherent slices, the total bunch duration is typically much longer than the coherence time. 
Likewise, a typical coherence slice with the above coherence time has a bandwidth of more than $\sigma_{\omega,0} = \sqrt{\pi}/\tau_{c,0}\gtrsim 3\,\mathrm{eV}/\hbar$, or $7\unit{eV}$ FWHM energy bandwidth $\Delta E$ ($\approx 0.1\%$), respectively~\cite{Schneidmiller2011}. 
Self seeded beams on the other hand have a much smaller bandwidth and longer coherence time. 
Typical values are less than $0.5\unit{eV}$ bandwidth, e.g. $\sigma_{E,0}=0.3\unit{eV}$ ($\Delta E=0.7\unit{eV}$ FWHM) for $17\unit{eV}$ photon energy ($0.004\%$) was achieved EuropeanXFEL. 
For $8\unit{keV}$, this would correspond to $\sigma_{E,0}\approx 0.14\unit{eV}$ ($\Delta E\approx 0.33\unit{eV}$ FWHM). 

To calculate the signal obtained under a certain angle $q$ when a coherent slice with central photon energy $\hbar\omega_0$, wave number $k_0$, and finite coherence time $\tau_{c,0}$ interferes with itself behind a warm plasma with scattering density $h\left(x\right)$ ($h$ being the product of the electron density and electron radius), we first start out with the well known equation for thin samples and in Born approximation and a fully coherent beam (i.e. monochromatic beam). 
Then the intensity is given by the absolute square $I_\omega(q)=\tilde f(q) \tilde f(q)^*$ of the Fourier transform $\tilde{f}(\vec{q}) = \int h(\vec{x})\,e^{-i\,\vec{q}\,\vec{x}} d^3x$, which again assuming the observation direction lies in the $x$-$z$ plane (i.e. $q \cong q_x$) is $\tilde{f}(q) = \int f(x)\,e^{-i\,q\,x} dx$ with  $f(x)=\iint_{-\infty}^{\infty}h(\vec{x})\,dy dz$. 
We rewrite $I_\omega(q)$ as
\begin{equation}
\begin{aligned}
I_\omega(q)&=\iint_{-\infty}^{+\infty} f(x_1)\,f(x_2)\, e^{-iq(x_1-x_2)}\,dx_1\,dx_2.
\end{aligned}
\end{equation}
The intensity $I(q)$ for \emph{finite} spectral width is then given by the weighted integral over all frequency components $I_\omega(q)$ where $f$ is weighted by the normalized spectral amplitude $s(\omega)=\sqrt{S(\omega)}$. 
We define the power spectral density after scattering at $x_i$ and experiencing a shift of the energy $\delta_i$ and the wavepacket broadening $d_i$ as
as
\begin{equation}
S_i(\omega)=\exp\!\left[-\frac{(\omega-\left(\omega_0+\delta_i\right)^2}{2\,(\sigma_{\omega,0}+d_i)^2}\right]. 
\end{equation}
For notational convenience we identify
\begin{equation}
q\,(x_1-x_2)=k_0\xi\equiv \omega_0 \tau,
\label{eqn:def_qY}
\end{equation}
as the phase difference of the two photon paths through $x_1$ and $x_2$ in the following, where $\xi$, $\tau$ are the path length difference in space and time domain, respectively. 
Then one can finally write
\begin{equation}
\begin{aligned}
I(q)=&\iint_{-\infty}^{+\infty} f\left(x_1\right)\, f\left(x_2\right)\,g(\tau)\,dx_1\,dx_2.
\label{eqn:Iq}
\end{aligned}
\end{equation}
where
\begin{equation}
g(\tau)\equiv\frac{G_{12}(\tau)}{\sqrt{G_{11}(0)\,G_{22}(0)}}
\end{equation}
with the cros correlation $G_{ij}(\tau)=\langle E_i(t)\,E_j^*(t+\tau)\rangle$, 
\begin{align}
G_{ij}(\tau)= \int_{-\infty}^{+\infty}\sqrt{S_i(\omega)\,S^*_j(\omega)}\,e^{-i\omega_0 \tau}d\omega
\end{align}
is the first order time cross-correlation function between the X-ray electric field amplitude $E$, and $\langle g(\tau) \rangle$ is the

Eqn.~\eqref{eqn:Iq} can be interpreted as the superposition of all possible paths of a photon, so what is left is to sum over all photons, i.e. up to a normalization constant that we will neglect here we can replace $g$ with the ensemble average $\langle g(\tau)\rangle$.
We will neglect small changes $d_i$ of the bandwidth in the following, as mentioned previously. 
I.e. with $\langle d_i^2 - \langle d_i\rangle^2\rangle \ll\sigma_{\omega,0}^2$ we can set, without loss of generality,  $\sigma_{\omega,0}+d_1=\sigma_{\omega,0}+d_2=\sigma_{\omega,0}+\langle d_i\rangle\equiv \sigma_{\omega,0}$ (see Appendix A and B for a rigorous justification). 
Then 
one can show that (see Appendix A) 
\begin{equation}
    g(\tau) =  \exp{\left[-\frac{(\delta_1-\delta_2)^2}{8\sigma_{\omega,0}^2}\,
              -i\,\omega_{eff}\,\tau\,
              -\frac{\sigma_{\omega,0}^2}{2}\,\tau^2\right]}\,
    \label{eqn:g}
\end{equation}
with the effective frequency
\begin{equation}
    \omega_{eff} = \omega_0 + \frac{\delta_1+\delta_2}{2}.
\end{equation}
Finally, we need to compute the ensemble averaged correlation function, which is given by $\langle g(\tau)\rangle = \int_{-\infty}^{+\infty}d\delta_1\,d\delta_2\,p(\delta_1)\,p(\delta_2)\,g(\tau)$. 
For normal distributed Doppler shifts the probability density $p$ is given by
\begin{equation}
    p(\delta_i)=\frac{1}{\sqrt{2\pi}\sigma_{\omega,D}}\,\exp{\left(\frac{\delta_i^2}{2\sigma_{\omega,D}^2}\right)} \,,\,i=1,2, 
\end{equation}
as they are expected for non-relativistic thermal plasmas, one obtains (see Appendix B)
\begin{equation}
\begin{aligned}
\langle g(\tau)\rangle &= 
              \sqrt{\frac{2\sigma_{\omega,0}^2}{2\sigma_{\omega,0}^2+\sigma_{\omega,D}^2}}\,
              \exp{\left(-i\,\omega_0\,\tau\,
              -\frac{\pi}{2}\,\frac{\tau^2}{{\tau_c}^2}\right)}\,
\label{eqn:avg_g}
\end{aligned}
\end{equation}
with the coherence time
\begin{equation}
\tau_c(\sigma_{\omega,D}) \equiv \int_{-\infty}^{+\infty}\left|g(\tau)\right|^2\,d\tau=\frac{\sqrt{\pi}}{{\sigma_{\omega}}}, 
\end{equation}
where ${\sigma_{\omega}}=\sqrt{\sigma_{\omega,0}^2+\sigma_{\omega,D}^2/2}$. 

Finally, we insert Eqn.~\eqref{eqn:avg_g} into \eqref{eqn:Iq} and replace $\omega_0\,\tau=q\,(x_1-x_2)$ (Eqn.~\eqref{eqn:def_qY}) to obtain
\begin{equation}
\begin{aligned}
I(q)=&a\,\iint_{-\infty}^{+\infty} f\left(x_1\right) f\left(x_2\right)\times \\
& \exp{\left[-\frac{\left(x_1-x_2\right)^2}{2\,L^2}+i\,q\,\left(x_1-x_2\right)\right]}\,dx_1\,dx_2. 
\label{eqn:I_broadband}
\end{aligned}
\end{equation}
with
\begin{equation}
    a=\frac{\tau_c}{\tau_{c,0}}=\frac{\sigma_{\omega,0}}{\sigma_{\omega}}=\sqrt{\frac{\sigma_{\omega,0}^2}{\sigma_{\omega,0}^2+\sigma_{\omega,D}^2/2}}
    \label{eqn:a}
\end{equation}
and
\begin{equation}
    {L}=\sqrt{\frac{1}{\pi}}\frac{k_0}{q}\,c\,\tau_c = \frac{\omega_0}{q\,{\sigma_{\omega}}}.
    \label{eqn:L}
\end{equation}

Before we proceed to two specific examples, let us make two remarks on equation~\eqref{eqn:I_broadband}. \\
First, we note that Eqn.~\eqref{eqn:I_broadband} is the intuitive and well known formula of visibility reduction of broadband beams [https://en.wikipedia.org/wiki/Higher\_order\_coherence] where $\sigma_{\omega,D}/2$ is the increase in bandwidth expected for a random, Gaussian process with variance $\sigma_{\omega,D}^2$, and with bandwidths adding quadratically;  concomitant with a reduction of the coherence time. 
While in a coherent wave packet all photons by definition have the same central energy, each scattering event changes the central frequency of a photon. 
Obviously, if multiple wave packets with different central energy interfere, the field's spectral width increases to the square root of the square sum of the width of the wavepackages and the width of the distribution of the central energies.
The formerly fully coherent slice of the XFEL pulse can hence be thought of rearranging into several shorter coherent slices which are incoherent to each other, which intuitively gives the reduction factor $\frac{\sigma_{\omega,0}}{\sigma_{\omega}}$ (Eqn.~\eqref{eqn:a}). \\
Secondly, it is evident that the broader the initial spectral width, the less significant role does the Doppler broadening play. 
Hence, the change induced by a warm plasma on the scattering pattern will be most important for seeded beams, and less for the more broadband SASE beams. 
Also, note that the coherence time and length are functions of scattering angle, cf. Eqn.~\eqref{eqn:doppler_shift}. 



\section{Examples}
\label{sec:Examples}

\begin{figure}
    \centering
    \includegraphics[width=\linewidth]{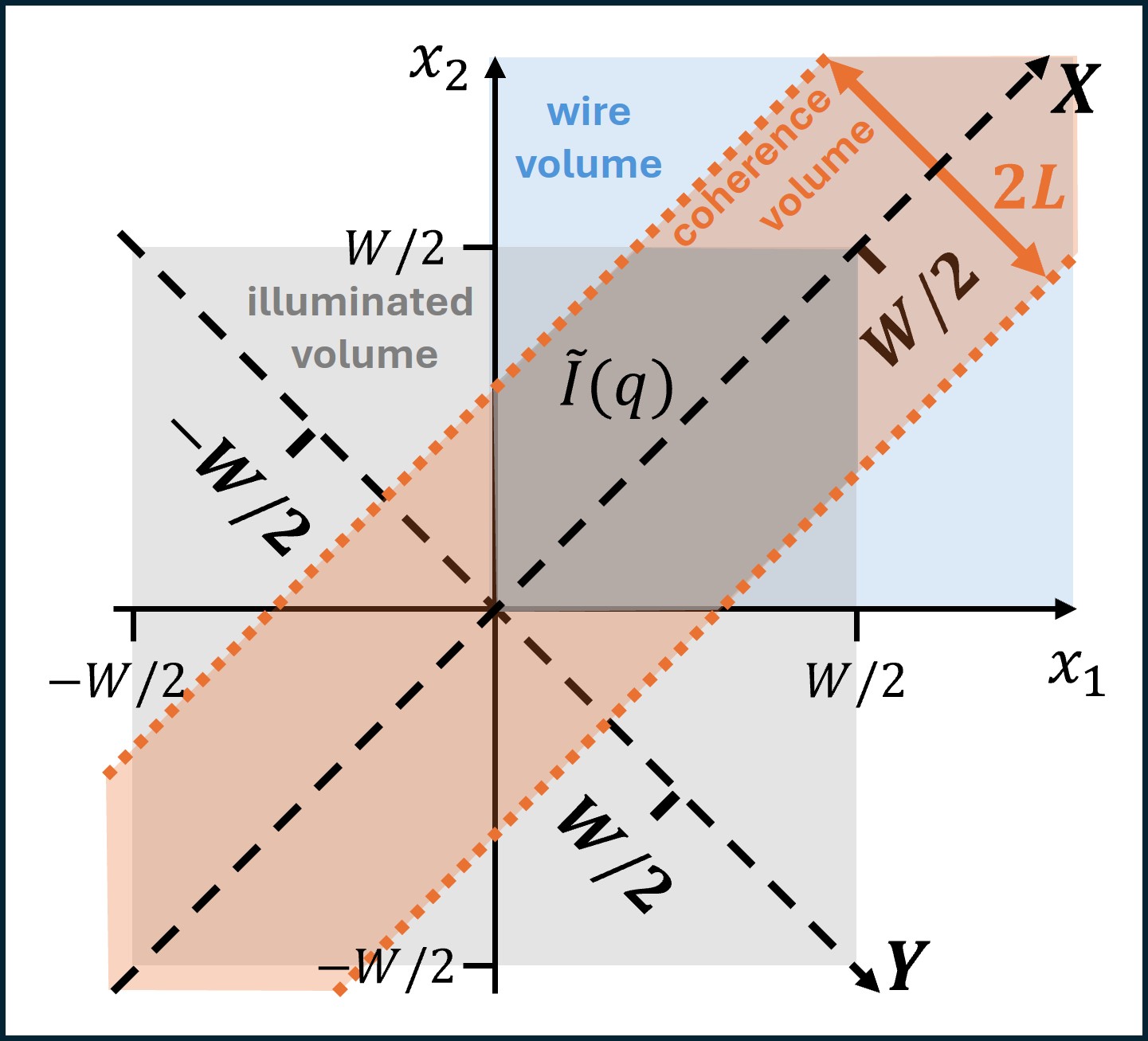}
    \caption{Visualizing the coordinate transform from $x_1,y_1$ to $X,Y$ and the integrand $M(X,Y)$ (Eqn.~\eqref{eqn:integration}). Light gray: illuminated area; orange: coherence area; blue: area filled by the wire; dark gray: area $I(q)$. }
    \label{fig:integrand}
\end{figure}

\subsection{Step function}
In the following we will examine the effects of Eqn.~\eqref{eqn:I_broadband} on the scattering signal for the two important cases of step targets and gratings. 

Let us assume the following step function type target with edge sharpness \(\sigma\), irradiated by a centered beam of width \(W\) which for simplicity we also model width an error function of width \(\Sigma\)
\begin{equation}
f(x)=\Biggl[\operatorname{erf}\!\left(\frac{x}{\sqrt{2}\,\sigma}\right) - \operatorname{erf}\!\left(\frac{x-\frac{W}{2}}{\sqrt{2}\,\Sigma}\right)\Biggr]. 
\label{eqn:step}
\end{equation}
This resembles for example the density of a tangentially irradiated wire with rectangular cross section. 
For small values of $\sigma\ll W$ and large \(q\) values $q\gg1/\sigma$, the scattering intensity for fully coherent beams, averaged over small oscillations in $q$, is then
\begin{equation}
    \begin{aligned}
        I(q)=&\left|\int_{-\infty}^{\infty} f(x)\,e^{-i\,q\,x}\,dx\right|^2\\
        =&\frac{1}{q^2}\Bigl(e^{-q^2\,\sigma^2} + e^{-q^2\,\Sigma^2} - \\
        &2\cos{\frac{W\,q}{2}}\,e^{-\frac{q^2}{2}(\sigma^2+\Sigma^2)}\Bigr). 
        \label{eqn:I_step}
    \end{aligned}
\end{equation}
This is essentially the well-known result for scattering from an edge of width $\sigma$ and another one with width $\Sigma$, plus a cross-term describing the interference between the two; for large values of $\Sigma \gg \sigma$ the scattering from the illumination outer edge is suppressed and we recover the single interface scattering $I(q)=q^{-2}\,\exp{\left(-q^2\,\sigma^2\right)}$; in the other limit of $\sigma=\Sigma=0$ the intensity gets fully modulated $$I(q)=\frac{4}{q_0^2}\sin^2{\frac{W\,q}{2}}.$$

We now want to study the influence of a finite bandwidth and Doppler shifts. 
We start by rewriting Eqn.~\eqref{eqn:avg_g} where we use $f(x)$ from Eqn.~\eqref{eqn:step}. 
Before looking at the general case, it is instructive to first study the analytically easily integrable approximation in the latter case of both a step-like wire surface and step-like irradiation function, and also binary coherence $\exp{\left(-Y^2/2{L}^2\right)}\longrightarrow \Theta({L} - |Y|)$. 
With these approximations we can write Eqn.~\eqref{eqn:avg_g} as
$I(q)=a\iint_{-\infty}^{\infty} M(X,Y) \,e^{-iq_0Y}\,dX\,dY$ with 
\begin{widetext}
    \begin{equation}
        \label{eqn:integration}
        M(X,Y) = \Theta\!\left(X+\frac{Y}{2}\right)
        \Theta\!\left(X-\frac{Y}{2}\right)
        \Theta\!\left({L}-|Y|\right)
        \Theta\!\left(\frac{W}{2}-\left|X+\frac{Y}{2}\right|\right)
        \Theta\!\left(\frac{W}{2}-\left|X-\frac{Y}{2}\right|\right)
    \end{equation}
\end{widetext}
The integrand $M$ has a simple geometric form (see Fig.~\ref{fig:integrand}): it can easily be integrated over \(X\) to yield
\begin{equation}
I(q)=a\left(\frac{W}{2}-|Y|\right)\,\Theta\!\left({L}-|Y|\right)\,\Theta\!\left(\frac{W}{2}-|Y|\right).
\end{equation}
Computing its Fourier transform in $Y$ we finally obtain for \(q\neq0\)
\begin{equation}
I(q)=a\,\left(\frac{\left(\frac{W}{2}-\frac{L_w}{2}\right)\sin(\frac{L_w}{2}\,q_0)}{q} + 4\frac{\sin^2(L_w\,q_0)}{q^2}\right), 
\label{eqn:grating}
\end{equation}
where $L_w$ is the smaller of ${L}$ or $W/2$. 
For finite coherence length ${L}<W/2$, the interference term oscillates slower than for a fully coherent beam, as the coherence length now limits the effective slit aperture instead of the larger illumination. 
\begin{figure}
    \centering
    \includegraphics[width=\linewidth]{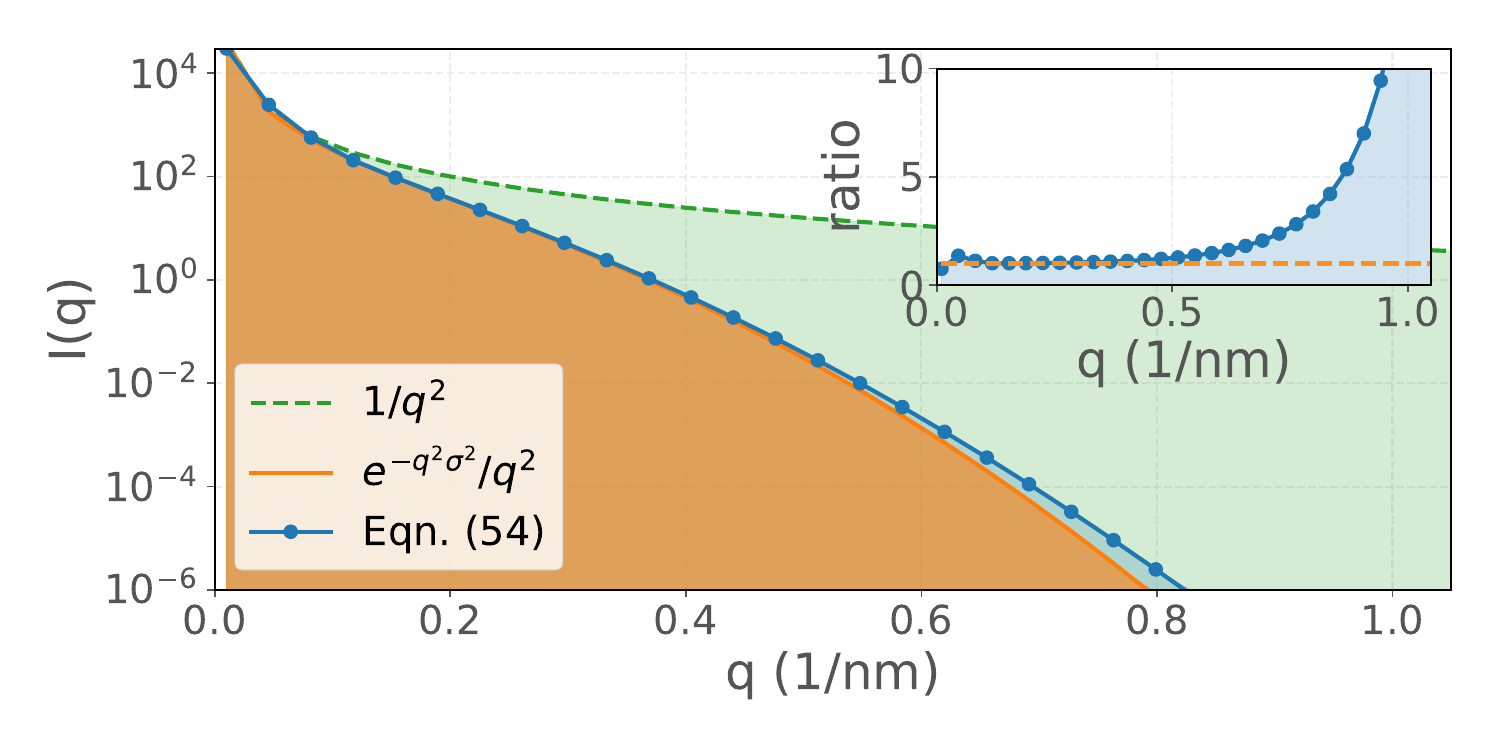}
    \caption{Effect of finite coherence on wire scattering on the example of $\sigma = 5\,\mathrm{nm}$. Here, we neglect the Doppler effect to keep $a=1$. $E_0$ is set to $8\unit{keV}$. Note that the energy spread is extremely large ($\Delta E = 400\unit{eV},\,\Delta E/E_0=5\%$) in order to emphasize the effective intensity increase. 
    \label{fig:wire_increase}}
\end{figure}

\begin{figure*}[t]
    \centering
    \includegraphics[width=1\linewidth]{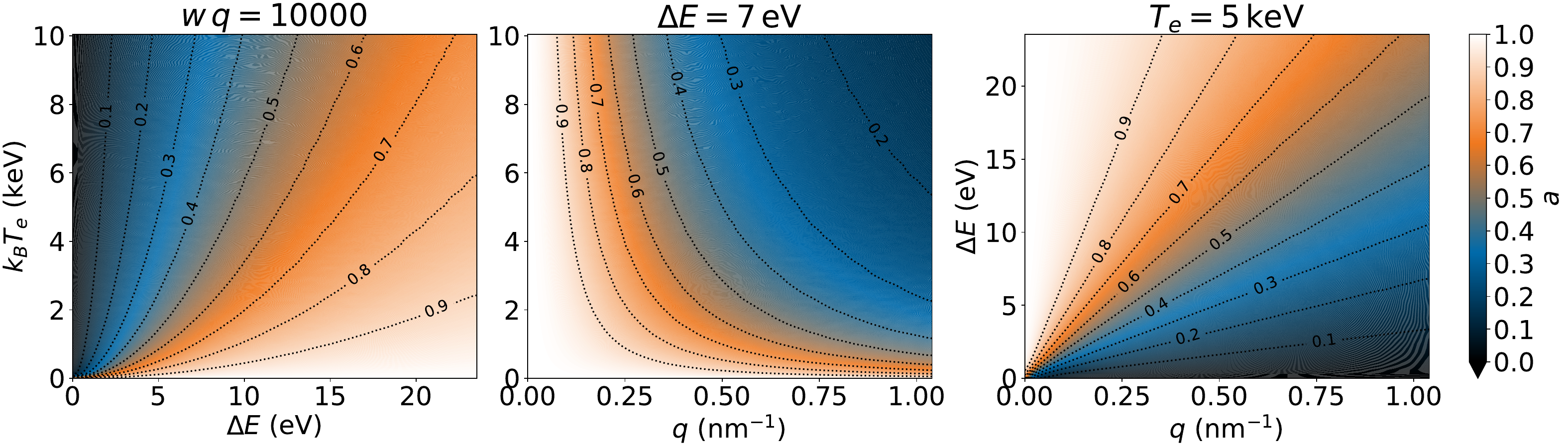}
    \caption{Reduction of scattering signal $I\propto a=\sigma_{\omega,0}/\sigma_\omega$ of a step-target (cp. Eqn.~\eqref{eqn:grating}), or peak integrated fluence $\int_{-\infty}^{+\infty}
    I_l(k)dk/\int_{-\infty}^{+\infty}
    I_{l,\omega_0}|_{T=0}dk=a$ of a grating target (cp Eqn.~\eqref{eqn:normalization}, in dependence of the three parameters electron temperature $T$, X-ray bandwidth FWHM $\Delta E = 2\sqrt{2\ln{2}}\,\sigma_{E,0}$, and wave number $q$. The bandwidth was chosen representative for SASE beams. The respective plots for parameter ranges relevant for seeded beams can be found in Appendix D for reference. 
    \label{fig:wires}}
\end{figure*}
This also causes the beam to diffract more, hence the intensity at $q_0\ne 0$ \emph{increases} by virtue of the first term in brackets. 
Hence, we find the opposite effect than in gratings, where a finite beam energy bandwidth caused a \emph{decrease} of intensity due to cutting coherent superposition of distant grating ridges. 
For wires, the scattering signal is essentially due only do a single interface, hence this cut is not effective here. \\
However, for most practical purposes, in small angle approximation it is ${L}>W/2$. 
Then the first term in brackets vanishes, i.e. the dependence on the bandwidth becomes negligible, and we recover the usual coherent sum of the scattering signal of both the wire density step and illumination edge given by Eqn.~\eqref{eqn:I_step} with $\sigma=\Sigma=0$, reduced by the partial loss of coherence expressed by the factor $a$. 
For example, for $E_0=8\unit{keV}$ photons, we see a distinct increase in scattering signal at $q_0=0.5/\mathrm{nm}$ only for $\Delta E\gtrsim 400\unit{eV}$. 
Hence from this simple analysis one finds that the finite bandwidth alone does not influence the scattering pattern of wires significantly, Doppler shifts however reduce the signal proportional to $a$. 

We now switch to the full description without assuming $\sigma=\Sigma=0$, i.e. we solve the full integral \eqref{eqn:avg_g} with $f$ from \eqref{eqn:step}. 
Since there exists no simple closed form solution of this integral, we have to solve it numerically. 
The numerical solutions support the main findings from the analytical simplified model. 
The leading effect of finite bandwidth and Doppler shifts is coming from $a$, see Fig.~\ref{fig:wires}. 
In Fig.~\ref{fig:wire_increase} we illustrate the possibility for an increasing intensity for an artificial extreme case. 

In Appendix C we demonstrate how the XFEL bandwidth and Doppler broadening disturb the reconstruction of the edge sharpness $\sigma$. For hard X-ray radiation, especially for sharp steps ($\sigma_0 = 1\unit{nm}$) and small X-ray beam bandwidths as in the cases of seeded XFELs the Doppler effect can be seen to completely render the smoothness extraction impossible without knowledge of the plasma temperature, even for low temperatures of few tens of eV. 
On the other hand, for smoother steps (e.g. $\sigma=3\unit{nm}$ and SASE-bandwidth the fitting error is $50\%$ or less even for high temperatures of few $\mathrm{keV}$. 

\subsection{Gratings}
Let us now assume a grating of size \(w\), pitch \(d\), and ridge steepness \(\sigma \ll d\). 
The illumination is assumed to be homogeneous and the number of grating ridges $N$ is much greater than $1$. 
Then the object function can be expressed as a Fourier series
\begin{equation}
    f(x)=\sum_{n=-\infty}^{\infty} c_n\, e^{i2\pi n x/d}.
    \label{eqn:fourier_series}
\end{equation}
Without loss of generality, we may assume symmetry $c_n=c^*_{-n}$. 
This we insert into Eqn.~\eqref{eqn:I_broadband} and change variables 
\begin{equation}
X=\frac{x_1+x_2}{2},\quad Y=x_1-x_2,
\end{equation}
such that
\begin{equation}
x_1=X+\frac{Y}{2},\quad x_2=X-\frac{Y}{2}, \quad dx_1\,dx_2=dX\,dY. 
\end{equation}
Integration along $Y$ is limited to the range $\left[\min{[x_1-x_2]},\,\max{[x_1-x_2]}\right]=[-w,+w]$. 
The limits for the integral along $X$ depend on $Y$, $\left[-(w-|Y|)/2,(w-|Y|)/2\right]$. 
We therefore have to solve the integral
\begin{equation}
\begin{aligned}
    I(q)\simeq a\,\sum_{n,m} c_n c_m &\int_{-w}^{w} e^{-\frac{Y^2}{2{L}^2}-i q Y+i\pi\frac{(n-m)Y}{d}} \\
    &\times\left[\int_{-\frac{w-|Y|}{2}}^{\frac{w-|Y|}{2}} e^{i2\pi\frac{(n+m)X}{d}}\,dX\right] dY.
\end{aligned}
\label{eqn:Iq_grating}
\end{equation}

The oscillatory terms in $X$ cancel out each other for $w\gg d$ and the integral over $X$ is governed by the non-oscillatory terms with $n+m=0$. 
This can easily be seen by comparing the integrals for $n+m=0$ and with those for  $n+m\ne 0$.
The former is simply given by $w-|Y|$ which is proportional to $w$, while the latter are $\frac{d}{\pi\left(n+m\right)}\sin{\left[\pi\left(n+m\right)\left(\frac{w-|Y|}{d}\right)\right]}<\frac{d}{\pi\left(n+m\right)}$, i.e. less than proportional to $d\ll w$. 

In the following we assume $w\gg d$, i.e. neglect side-lobes. 
Substituting the prefactors in the terms linear to $Y$ in the exponent of the $Y$-integral with
\begin{equation}
    k = k_n=\frac{2\pi\,n}{d}-q
\end{equation}
Eqn.~\eqref{eqn:Iq_grating} becomes
\begin{equation}
    \begin{aligned}
        I(q) \simeq &\,a\,\sum_n \left|c_n\right|^2 \int_{-w}^{w} \left(w - |Y|\right) e^{i k Y} e^{-Y^2/(2{L}^2)} \, dY \\
        = &\,a\,\sum_n \left|c_n\right|^2 \left[ \int_{-w}^{0} (w + Y) e^{i k Y} e^{-Y^2/(2{L}^2)} \, dY \right.\\
        &\left.+ \int_{0}^{w} (w - Y) e^{i k Y} e^{-Y^2/(2{L}^2)} \, dY \right] \\
        = &\, 2\,a\,\sum_n \left|c_n\right|^2\,\Re\left(J_n\right).
    \end{aligned}
    \label{eqn:fullIq_grating}
\end{equation}
The integral 
\begin{equation}
    J_n = \int_{0}^{w}\left(w-Y\right)\, e^{i\,k\,Y} e^{-Y^2/(2{L}^2)}\,dY
    \label{eqn:Jn_integral}
\end{equation}
can be separated in two standard integrals 
\begin{equation}
    J_n = w\,K_1-K_2
    \label{eqn:Jn}
\end{equation}
with
\begin{equation}
    \begin{aligned}
        K_1 &= \int_{0}^{w} e^{i k Y - \frac{Y^2}{2 {L}^2}}\, dY \\
        &= \sqrt{\frac{\pi}{2}}\,L\, e^{-\frac{k^2 {L}^2}{2}}\left[\mathrm{erf}\left(\frac{w - i k {L}^2}{\sqrt{2} {L}}\right) \right.\\
        &\quad\quad\quad\quad\quad\quad\quad\quad\,\,\left.+\,\mathrm{erf}\left(i\frac{k {L}}{\sqrt{2}}\right)\right]
        \label{eqn:J1}
    \end{aligned}
\end{equation}
and
\begin{equation}
    \begin{aligned}
        K_2 &= \int_{0}^{w} Y\,e^{i\,k\,Y} e^{-Y^2/(2{L}^2)}\,dY\\
        &={L}^2\left[1-e^{-w^2/2{L}^2+i\,k\,w}\right] + i\,k\,{L}^2\,K_1.
        \label{eqn:J2}
    \end{aligned}
\end{equation}
This is the most general closed form solution. 

It can easily be shown that $I(q)$ is still peaked around $k=0$, i.e. $q_l\approx 2\pi\,l/d$. 
In the following we will derive first how the peak height and peak width change for warm plasmas, i.e. for changes of bandwidth and coherence. 
This we will use then to derive the influence onto experimental measurements of the peak signal in the two limiting cases of high resolution pixel detectors that can resolve the peak shape, and low resolution where only the integrated fluence is measured. 

\subsubsection{Peak height}
\begin{figure*}
    \centering
    \includegraphics[width=\linewidth]{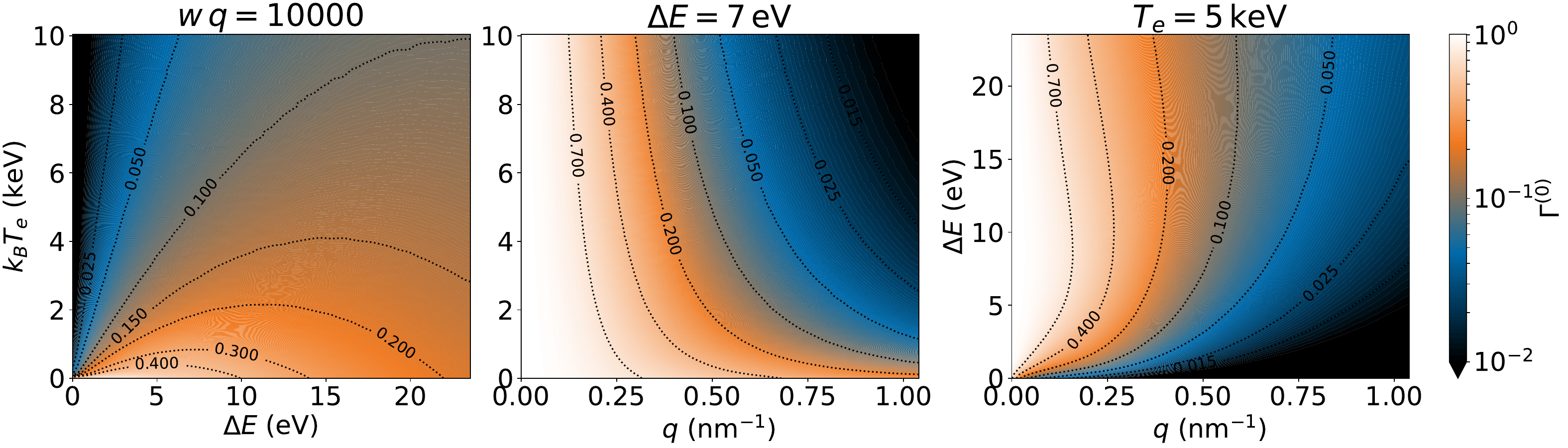}
    \caption{Effect of finite coherence and Doppler shift on grating peak height $\Gamma(0)$ (Eqn.~\eqref{eqn:grating_amplitude} with $a,L$ from Eqns.~\eqref{eqn:a},~\eqref{eqn:L}) in dependence of the three parameters electron temperature $T$, X-ray bandwidth FWHM $\Delta E = 2\sqrt{2\ln{2}}\,\sigma_{E,0}$, and wave number $q$. The bandwidth was chosen representative for SASE beams. The respective plots for parameter ranges relevant for seeded beams can be found in Appendix D for reference
    \label{fig:gratings}}
\end{figure*}

The signal around one peak is given by one summand of Eqn.~\eqref{eqn:fullIq_grating}. 
Hence the height of peak with index $l$ normalized to its height for a monochromatic X-ray beam and cold plasma $I_0=I_{l,\omega_0}|_{T=0}(k=0)=\left|c_l\right|^2\,w^2$ can be expressed by 
\begin{equation}
    \Gamma(k)\equiv \frac{I_l(k)}{I_0}=\frac{2\,a}{w^2}\,\Re{\left(J_l(k)\right)}
    \label{eqn:Gamma}
\end{equation}
where $I_l(k)=I(k)|_{\tilde c_i=\tilde c_l\delta_n}$, i.e. all $c_i$ for $i\ne l$ are set to zero. 
For $k=0$ the integrals greatly simplify, and the peak height reduction factor can be compactly written as
\begin{equation}
\begin{aligned}
\Gamma(0) = &\,2\,a\,\frac{L}{w}
\Biggl[
\sqrt{\frac{\pi}{2}}\;\operatorname{erf}\!\Bigl(\frac{w}{\sqrt{2}\,L}\Bigr)\\
&\qquad\;-\;\frac{L}{w}\;\Bigl(1 - \exp\bigl(-\tfrac{w^2}{2\,L^2}\bigr)\Bigr)
\Biggr].
\label{eqn:grating_amplitude}
\end{aligned}
\end{equation}
This effect of the finite bandwidth (i.e. coherence) and Doppler shifts in warm plasmas on reducing the peak heights (i.e. peak visibility) is visualized for some exemplary parameter ranges in Fig.~\ref{fig:gratings}.

\subsubsection{Peak width}
\label{sec:ExamplesGratingsPeakwidth}
Expanding the integrals \eqref{eqn:J1} and \eqref{eqn:J2} around $k=0$ shows that at the same time as the peak height reduces the shape of the peaks becomes Gaussian, whose width increases with smaller values of ${L}$ (i.e. larger $q$ and larger bandwidth $\sigma_\omega$): 
The Taylor expansion of the real part of Eqn.~\eqref{eqn:Gamma} around $k=0$ gives
\begin{equation}
    \frac{\Gamma(k)}{\Gamma^{(0)}}=1- \frac{k^2}{2}\alpha^2+\frac{k^4}{8}\beta^4+\mathcal{O}(k^6)
    \label{eqn:expansionGamma}
\end{equation}
with 
\begin{equation}
    \begin{aligned}
        \alpha^2 = \frac{2\,a\,L^2}{\Gamma^{(0)}} \biggl[ &
        \sqrt{\frac{\pi}{2}}\,\frac{L}{w} \operatorname{erf}\left(\frac{w}{\sqrt{2}L}\right) \\
        &-2\frac{{L}^2}{w^2 } \left(1 - e^{-w^2/(2L^2)} \right) \biggr]
    \end{aligned}
\end{equation}
and 
\begin{equation}
    \begin{aligned}
    \beta^4 = \frac{2\,a\,L^{\prime\,4}}{\Gamma^{(0)}} \biggl[ &
\sqrt{\frac{\pi}{2}}\frac{L}{w}\, \operatorname{erf}\left(\frac{w}{\sqrt{2}L} \right) \\
        & - \frac{8{L}^2}{3\,w^2} \biggl(1 - e^{-w^2/(2L^2)}\\
        &\quad\times\left[1+\frac{w^2}{2L^2}+\frac{w^4}{8L^{\prime\,4}}\right] \biggr)\\
        &    -\frac{1}{3}\,e^{-w^2/(2L^2)}\left(1-\frac{w^2}{{L}^2}\right) \biggr]
    \end{aligned}
\end{equation}
which for $L\ll w$ simplifies to 
\begin{equation}
    \begin{aligned}
        \alpha &= L\left(1-\sqrt{\frac{2}{\pi}}\frac{L}{w}\right)+\mathcal{O}\left[\left(\frac{L}{w}\right)^2\right], \\
        \beta &= L\left(1-\frac{2}{3}\sqrt{\frac{2}{\pi}}\frac{L}{w}\right)+\mathcal{O}\left[\left(\frac{L}{w}\right)^2\right].
    \end{aligned}
\end{equation}
Neglecting terms of $L/w$, the expansion \eqref{eqn:expansionGamma} with $\alpha=\beta=L$ is that of a Gaussian with width $1/L$.
The integral over k trivially yields
\begin{equation}
I_{l,int}\equiv\int_{-\infty}^{\infty} I_l(k)\, dk =  2\pi\,\left|c_l\right|^2\,a\,w, 
\label{eqn:normalization}
\end{equation}
which is independent of \({L}\) as expected from Parseval's theorem. 
The overall integrated scattering signal around the main peaks reduces due to the loss of coherence through Doppler shifts only by virtue of the pre-factor $a$. 
In fact, Eqn.~\eqref{eqn:normalization} is true exactly for arbitrary values of $k$, $L$ and $w$, as can quickly be shown by integrating one of the summands of Eqn.~\eqref{eqn:fullIq_grating} with \eqref{eqn:Jn_integral} over $k$ around $k=0$.
Interchanging the order of integration results in 
\[
\begin{aligned}
&\int_{-\infty}^{\infty} I_l\, dk 
=\\
&\quad\quad a\,\left|c_l\right|^2\,\int_{-w}^{w} \Bigl( w - |y| \Bigr) e^{-\frac{y^2}{2{L}^2}} \left[ \int_{-\infty}^{\infty} e^{ik y}\, dk \right] dy,
\end{aligned}
\]
where the \(k\)–integral can be recognized as the Fourier representation of the Dirac delta function
\[
\int_{-\infty}^{\infty} e^{i k y}\, dk = 2\pi\,\delta(y).
\]
Thus, the expression becomes
\begin{equation}
I_{l,int}=\int_{-\infty}^{\infty} I_l(k)\, dk =  2\pi\,\left|c_l\right|^2\,a\,w,
\label{eqn:normalizationFull}
\end{equation}
which is the same as \eqref{eqn:normalization}. 
The reduction in integrated fluence is given by $a$, which is shown in Fig.~\ref{fig:wires} for some relevant parameter ranges. 

\subsubsection{Peak signal measurement}
\label{subsubsec:eff}
The practical relevant measure often is the integrated signal over a peak on the detector. In practice, the signal can only be measured in a region $[k_-,k_+]$ where it is visible, i.e. where the signal to noise ratio $SNR > 1$. 
Hence the practical measure of the influence of finite bandwidth and Doppler shifts is not given by $a$ from Eqn.~\eqref{eqn:normalizationFull} by integration over the full peak but rather the ratio
\begin{equation}
    a_{eff}(q_l)=\frac{\int_{k_-}^{k_+} I_{coh}(k)\, dk}{\int_{-\infty}^{\infty} I_{l,\omega_0}|_{T=0}(k)\, dk}
    \label{eqn:aeff}
\end{equation}
the denominator is simply the normalization given by Eqn.~\eqref{eqn:normalizationFull} with $a=1$. 
We now need to obtain the integration limits $k_\pm$. 
The measured signal generally consists of the sum of the coherent scattering signal $N_{coh}\cong I_{coh}(q)\cdot\Delta q$ ($\Delta q$ being the detector pixel size) that we treated above, and the incoherent scattering signal $N_{incoh} = b \approx const.$ which may include also external sources such as bremsstrahlung, detector noise etc.  
The coherent signal can usually be obtained by subtracting the constant background from the measured signal. 
The noise level of the thus calculated coherent intensity can be approximated by Poisson statistics using the measured photon numbers $\left(N_{coh}+N_{incoh}\right)^{1/2}$, hence the signal to noise level at each peak with index $l$ is given by [Hughes \& Hase, 2009]
\begin{equation}
    \begin{aligned}
        \mathrm{SNR} &= \frac{N_{coh}(k)}{\sqrt{N_{coh}(k)+N_{incoh}}}. 
    \end{aligned}
\end{equation}
There are now two limiting cases: Either the detector can resolve the peak width, i.e. $L\,\Delta q \ll 1$, then for $L \ll w$ it is
\begin{equation}
    N_{coh}(k) = N_l(k) \approx N_0\,\exp{\left[-k^2/2\Delta k^2\right]}
\end{equation}
where from $\Delta k = 1/L$ and  $N_0=\left|c_l\right|^2\,w^2\,\Gamma^{(0)}\,\Delta q\cong \sqrt{2\pi}\left|c_l\right|^2\,a\,w\,L\,\Delta q$ is proportional to the peak height, see Sec.~\ref{sec:ExamplesGratingsPeakwidth};  
or in the opposite case when $L\,\Delta q \gg 1$ the peak width on the detector is given by the detector point spread function which one often may approximate as a Gaussian with width $\Delta k\equiv \Delta Q_D$ and height $N_0\cong\left(2\pi\right)^{1/2}\,|c_l|^2\,a\,w\,\Delta q/\Delta Q_D$ (assuming $100\%$ quantum efficiency). 

In both cases we can insert the Gaussian shape function into Eqn.~\eqref{eqn:aeff}
\begin{equation}
    a_{eff}(q_l)=a\,\sqrt{\frac{1}{2\pi}}\frac{1}{\Delta k}\,\int_{k_-}^{k_+}\,\exp{\left(-k^2/2\Delta k^2\right)}\,dk.
\end{equation}
and then find for the integration bounds
\begin{equation}
k_{\pm} = \pm \Delta k\,\sqrt{2\,\ln\!\left[\frac{2\,N_0}{1 + \sqrt{1 + 4\,N_{incoh}}}\right]}.
\end{equation}
Then one finally obtains
\begin{equation}
    a_{eff}(q_l)=a\,\operatorname{erf}\left(\sqrt{\ln{\left[\frac{2\,N_0}{1+\sqrt{1+4\,N_{incoh}}}\right]}}\right). 
    \label{eqn:F}
\end{equation}
This is the effective measured reduction of the grating scattering peaks for warm plasmas as compared to the cold plasmas and including the limited peak visibilty against the incoherent signal. 
Note that in the case when the detector can resolve the peak width, $L\,\Delta q \ll 1$, $N_0$ is proportional to $\Gamma^{(0)}$, while in the opposite case it is $N_0\propto a$. 

In Appendix C we show some examples for both cases. 
There we demonstrate how the XFEL bandwidth, plasma temperature, signal level and background noise contribute to measurement errors for example of the grating sharpness $\sigma$. 
Especially in the case of self-seeded XFEL beams plasma temperatures of less than$100\unit{eV}$ can already lead to a significant error in extraction of $\sigma$, while for SASE beams only temperatures in excess of a few $100\unit{eV}$ lead to similar errors. 

\section{Conclusions}
In this paper we studied the influence of temperature effects on small angle scattering for the important examples of a grating and wire. 
The Doppler effect in heated plasmas leads to a net broadening of the energy bandwidth proportional to the photon scattering angle. 
As soon as the broadening gets into the range of the beam bandwith, which is generally the case for larger angles, this leads to a reduction of the coherent interference scattering signal. 
While these effects are generally minor for most cases of SASE XFEL beams and sub-keV plasma temperature, for seeded pulses or higher temperatures the effects can become large. 
Then they have to be included in the reconstruction algorithms and can lead to significant deviation in the extracted geometric parameters such as the sharpness of grating ridges or a surface/interface. \\
In the other direction, one might be interested to use the Doppler effect on the scattering signal to quantitatively determine the plasma temperature. 
While this is in principle possible, in practice the correlation of the temperature and bandwidth effects with geometric effects through ablation, melt or compression needs to be resolved. 
This will require both background free measurements with high statistical quality (i.e. high count-rates), as well as measurements over a large q-range to be able to resolve the different quantitative $q$-dependencies. 
\begin{acknowledgments}
t.b.d.
\end{acknowledgments}



\section*{Author Contributions}
All authors reviewed the manuscript and contributed extensively to the discussions.

\section*{Competing interests}
All authors declare no competing interests.

\section*{References}
\bibliographystyle{sn-mathphys-num}
\bibliography{Mendeley}

\clearpage
\begin{widetext}
\section*{Appendix A}
To derive Eqn.~\eqref{eqn:g} we start with $g(\tau)=G_{12}(\tau)/\sqrt{G_{11}(0)\,G_{12}(0)}$. First, we write $G_{12}(\tau)$ explicitly 
\begin{equation}
    \begin{aligned}
        G_{12}(\tau) &= \int_{-\infty}^{+\infty}\exp{\left[-\frac{(\omega-\left(\omega_0+\delta_1\right)^2}{4\,(\sigma_{\omega,0}+d_1)^2}\right]}\,\exp{\left[-\frac{(\omega-\left(\omega_0+\delta_2\right)^2}{4\,(\sigma_{\omega,0}+d_2)^2}\right]}\,\exp{\left[-i\omega \tau\right]}d\omega\\
        &\equiv \int_{-\infty}^{+\infty}\,\exp{\left[-\frac{(\omega-\omega_{1})^2}{4\,\sigma_{\omega,1}^2}-\frac{(\omega-\omega_{2})^2}{4\,\sigma_{\omega,2}^2}-i\omega \tau\right]}d\omega.  \\ 
    \end{aligned}
\end{equation}
We now expand the squares
\[
\begin{aligned}
-\frac{(\omega-\omega_{i})^2}{4\sigma_{i}^2} &= -\frac{1}{4\sigma_{i}^2}\left[\omega^2 - 2\omega_{i}\,\omega + \omega_{i}^2\right],\,\,i=1,2.
\end{aligned}
\]
This gives for the exponent
\[
-\left(\frac{1}{4\sigma_{\omega,1}^2}+\frac{1}{4\sigma_{\omega,2}^2}\right)\omega^2
+\left(\frac{\omega_{1}}{2\sigma_{\omega,1}^2}+\frac{\omega_{2}}{2\sigma_{\omega,2}^2}-i t\right)\omega
-\left(\frac{\omega_{1}^2}{4\sigma_{\omega,1}^2}+\frac{\omega_{2}^2}{4\sigma_{\omega,2}^2}\right).
\]
To write this in the standard Gaussian form, we define the following coefficients:
\[
A=\frac{1}{4\sigma_{\omega,1}^2}+\frac{1}{4\sigma_{\omega,2}^2}=\frac{\sigma_{\omega,1}^2+\sigma_{\omega,2}^2}{4\sigma_{\omega,1}^2\sigma_{\omega,2}^2},
\]
\[
B=\frac{\omega_{1}}{2\sigma_{\omega,1}^2}+\frac{\omega_{2}}{2\sigma_{\omega,2}^2}-i\tau,
\]
\[
C=-\left(\frac{\omega_{1}^2}{4\sigma_{\omega,1}^2}+\frac{\omega_{2}^2}{4\sigma_{\omega,2}^2}\right).
\]
Then we one can write
\[
G_{12}(t)=e^{\,C}\int_{-\infty}^{\infty}e^{-A\omega^2+B\omega}\,d\omega=e^{\,C}\,\sqrt{\frac{\pi}{A}}\,\exp\!\left(\frac{B^2}{4A}\right).
\]
We now substitute $A$, $B$ and $C$. 
Expanding \(B^2\) gives
\[
B^2=\left(\frac{\omega_{1}\sigma_{\omega,2}^2+\omega_{2}\sigma_{\omega,1}^2}{2\sigma_{\omega,1}^2\sigma_{\omega,2}^2}-it\right)^2
=\frac{(\omega_{1}\sigma_{\omega,2}^2+\omega_{2}\sigma_{\omega,1}^2)^2}{4\sigma_{\omega,1}^4\sigma_{\omega,2}^4}
- i\,\tau\,\frac{\omega_{1}\sigma_{\omega,2}^2+\omega_{2}\sigma_{\omega,1}^2}{\sigma_{\omega,1}^2\sigma_{\omega,2}^2}-t^2.
\]
Thus,
\begin{equation}
\frac{B^2}{4A}+C=-\left(\frac{\omega_{1}^2}{4\sigma_{\omega,1}^2}+\frac{\omega_{2}^2}{4\sigma_{\omega,2}^2}\right)
+\frac{(\omega_{1}\sigma_{\omega,2}^2+\omega_{2}\sigma_{\omega,1}^2)^2}{4\sigma_{\omega,1}^2\sigma_{\omega,2}^2(\sigma_{\omega,1}^2+\sigma_{\omega,2}^2)}
-i\,\tau\,\frac{\omega_{1}\sigma_{\omega,2}^2+\omega_{2}\sigma_{\omega,1}^2}{\sigma_{\omega,1}^2+\sigma_{\omega,2}^2}
-\frac{\sigma_{\omega,1}^2\sigma_{\omega,2}^2}{\sigma_{\omega,1}^2+\sigma_{\omega,2}^2}\,\tau^2.
\label{eqn:appendix_G12_almost_done}
\end{equation}
As the last step, we need to simplify the $\tau$-independent terms.
After some lengthy but straightforward algebra and the trivial result $G_{ii}(0) = \sqrt{2\pi}\sigma_{i}$ one finds
\begin{equation}
    \boxed{
        \begin{aligned}
        g(\tau)=\; &\frac{2\sigma_{\omega,1}\sigma_{\omega,2}}{\sigma_{\omega,1}^2+\sigma_{\omega,2}^2}\,\exp\Biggl[-\frac{(\omega_{1}-\omega_{2})^2}{4(\sigma_{\omega,1}^2+\sigma_{\omega,2}^2)}\\[1mm]
        &\quad{}-i\,\frac{\omega_{1}\sigma_{\omega,2}^2+\omega_{2}\sigma_{\omega,1}^2}{\sigma_{\omega,1}^2+\sigma_{\omega,2}^2}\,t
        -\frac{\sigma_{\omega,1}^2\sigma_{\omega,2}^2}{\sigma_{\omega,1}^2+\sigma_{\omega,2}^2}\,t^2\Biggr].
        \label{eqn:appendixA_result}
        \end{aligned}
    }
\end{equation}

For $\sigma_{\omega,1}\approx \sigma_{\omega,2}\equiv\sigma_{\omega}$ we finally obtain Eqn.~\eqref{eqn:g}.

\section*{Appendix B}
To obtain Eqn.~\eqref{eqn:avg_g}, we need to average Eqn.~\eqref{eqn:appendixA_result} over the energy shifts $\delta_i$ and bandwidth fluctuations $d_i$. 
The average over $d_i$ yields no closed form solution in general. 
We therefore assume $d_i\ll\sigma_{i}$. 
Likewise, for $\tau >> \tau_c$ it is trivial that $\lim_{\tau\longrightarrow \infty} g(\tau) = 0$ hence the interesting case is $\tau \ll \tau_c$ and we may therefor also adopt $d_i\,\tau \ll 1$ and develop Eqn.~\eqref{eqn:avg_g} to the leading order in $d_i/\sigma_{\omega}$ and $d_i\,\tau$. 
Expanding $g(\tau)$ to first order in the small parameters we can write $g(\tau)=g_0(\tau)\cdot g_1(\tau)$ 
with
\begin{equation}
    \begin{aligned}
        g_0(\tau)&=\exp\!\left[-\frac{(\delta_1-\delta_2)^2}{8\sigma_{\omega,0}^2}-i\frac{\delta_1+\delta_2}{2}\tau\right]\,\exp{\left(-i\,\omega_0\tau-\frac{\sigma_{\omega,0}^2}{2}\,\tau^2\right)}\\
        g_1(\tau) &\approx \, \exp\left[-\frac{(\omega_{1}-\omega_{2})^2}{8\sigma_{\omega,0}^2}\left(\frac{d_1+d_2}{\sigma_{\omega,0}}\right) -\,i\,\tau\,\frac{(\omega_{1}-\omega_{2})}{2\sigma_{\omega,0}} (d_2-d_1) -\frac{\tau^2}{2}\sigma_{\omega,0}\,(d_1+d_2)
    \right].
    \end{aligned}
\end{equation}
Hence, the average decouples into $\left<g(\tau)\right>_{\delta,d} = \left<g_0(\tau)\right>_\delta\cdot\left<g_1(\tau)\right>_d$
%
We first treat the first term, i.e. we first calculate the average of $g_0$ over small energy shifts. 
Only the first term, which we refer to a $K$ in the following, contains $\delta_i$,
\[\left\langle K \right\rangle_\delta \equiv \left\langle \exp\!\left[-\frac{(\delta_1-\delta_2)^2}{8\sigma_{\omega,0}^2}\right]\,\exp\!\left[-i\frac{\delta_1+\delta_2}{2}\tau\right]\right\rangle_\delta
=\iint_{-\infty}^{\infty} \frac{d\delta_1\,d\delta_2}{2\pi\,\sigma_{\omega,D}^2}\,
\exp\!\left[-\frac{\delta_1^2+\delta_2^2}{2\sigma_{\omega,D}^2}-\frac{(\delta_1-\delta_2)^2}{8\sigma_{\omega,0}^2}-i\frac{\delta_1+\delta_2}{2}\tau\right],
\]
where \(\delta_1\) and \(\delta_2\) are independent Gaussian random variables with variance \(\sigma_{\omega,D}^2\).
It is convenient to change variables to
\[
\delta_+ = \delta_1 + \delta_2,\quad \delta_- = \delta_1 - \delta_2,\quad
d\delta_1\,d\delta_2 = \frac{1}{2}\,d\delta_+\,d\delta_-.
\]
Then
\[
\delta_1^2+\delta_2^2 = \frac{1}{2}\left(\delta_+^2 + \delta_-^2\right).
\]
Thus, the integral becomes
\[
\left\langle K \right\rangle_\delta = \frac{1}{2\pi\,\sigma_{\omega,D}^2}\cdot\frac{1}{2}\int_{-\infty}^{\infty} d\delta_+ \int_{-\infty}^{\infty} d\delta_-\,
\exp\!\left[-\frac{\delta_+^2+\delta_-^2}{4\sigma_{\omega,D}^2}-\frac{\delta_-^2}{8\sigma_{\omega,0}^2}-i\frac{\delta_+}{2}\tau\right].
\]
Since the integrals decouple, we can write
\[
\left\langle K \right\rangle_\delta = \frac{1}{4\pi\,\sigma_{\omega,D}^2}\,
\left[\int_{-\infty}^{\infty} d\delta_+\,\exp\!\left(-\frac{\delta_+^2}{4\sigma_{\omega,D}^2}-i\frac{\delta_+\,\tau}{2}\right)\right]
\left[\int_{-\infty}^{\infty} d\delta_-\,\exp\!\left(-\frac{\delta_-^2}{4\sigma_{\omega,D}^2} - \frac{\delta_-^2}{8\sigma_{\omega,0}^2}\right)\right].
\]
Both integrals are standard Gaussian integrals after completing the squares in the exponents. 
The \(\delta_+\) integral is:
\[
\int_{-\infty}^{\infty} d\delta_+\,\exp\!\left(-\frac{\delta_+^2}{4\sigma_{\omega,D}^2}-i\frac{\delta_+\,\tau}{2}\right) = 2\sqrt{\pi}\,\sigma_{\omega,D}\,\exp\!\left(-\frac{\sigma_{\omega,D}^2\tau^2}{4}\right)
\]
The \(\delta_-\) integral is
\[
\int_{-\infty}^{\infty} d\delta_-\,\exp\!\left[-\frac{2\sigma_{\omega,0}^2+\sigma_{\omega,D}^2}{8\sigma_{\omega,D}^2\,\sigma_{\omega,0}^2}\delta_-^2\right]
=\sqrt{\frac{8\pi\,\sigma_{\omega,D}^2\,\sigma_{\omega,0}^2}{2\sigma_{\omega,0}^2+\sigma_{\omega,D}^2}}.
\]
Substituting the two integrals back into \(K\)
one readily obtains
\[
\boxed{
\left\langle K \right\rangle_\delta
=\sqrt{\frac{2\sigma_{\omega,0}^2}{2\sigma_{\omega,0}^2+\sigma_{\omega,D}^2}}\,\exp\!\left(-\frac{\sigma_{\omega,D}^2\tau^2}{4}\right)\,}
\]
and with $\left<g_0(\tau)\right> = K\,\exp{\left(-i\,\omega_0\tau-\sigma_{\omega,0}^2\,\tau^2/2\right)}$ (Eqn.~\eqref{eqn:g}) one obtains Eqn.~\eqref{eqn:avg_g}. 

Now we are left computing $\left<g_1(\tau)\right>_d$. We define $u=d_1+d_2$ with $\left<u\right>=0$, $\left<u^2\right>=2\sigma_{\Delta}$ for Gaussian distributed $d_i$ with width $\sigma_{\Delta}$. 
Using the identity $\left<\exp(k\,u)\right>=\exp\left(k^2 \sigma_{\omega,0}^2/2\right)$ which holds for any zero–mean Gaussian 
$u$ with variance $\sigma_{\omega,0}^2$. 
Hence,
\begin{equation}
    \left<g_1\left(\tau\right)\right>_d = \exp{\left(\frac{\sigma_{\omega,0}^2\,\sigma_{\Delta}^2\,\tau^4}{4}\right)}. 
\end{equation}
Since we assumed $\sigma_\Delta\,\tau\ll 1$, the exponential is a second order small correction to the $\exp{\left(-\sigma_{\omega,0}^2\,\tau^/2\right)}$ term in $\left<g_0\right>$ and hence can be neglected, $\langle g_1 \rangle_d \approx 1$, hence $g\cong g_0$. 

\section*{Appendix C}

The effect of finite bandwidth and Doppler broadening on the reduction of the grating peak height and integrated fluence can be used to characterize the plasma temperature. 
On the other hand, it may induce errors when trying to user the scattering pattern to retrieve the grating geometry since both effects on the scattering signal are correlated. 
In order to develop a quantitative understanding of the significance of these effects, we show the calculated peak shapes (Fig.~\ref{fig:peak_broadening}), and signal reduction together with the change of the best fits and the connected model parameters (Figs.~\ref{fig:fitting9}ff). 
\begin{figure}[!h]
    \centering
    \includegraphics[width=0.4\linewidth]{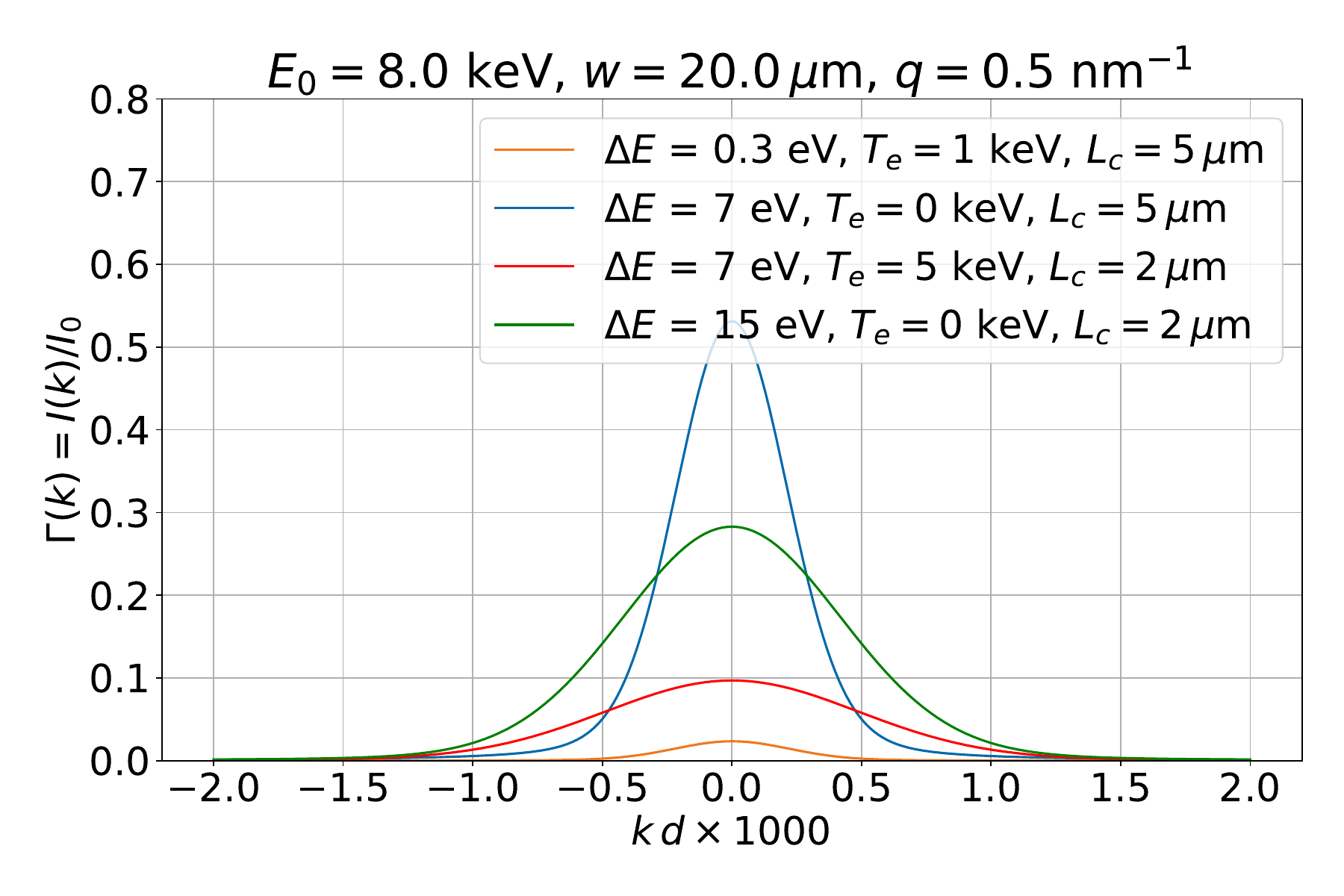}
    \caption{Calculated peak shapes for different combinations of bandwidth and plasma temperature. 
    \label{fig:peak_broadening}}
\end{figure}
Specifically, in Figs.~\ref{fig:fitting9}-\ref{fig:fitting8} we show the coherent signal as given by the model function $m = A\cdot \exp{\left(-q^2\,\sigma^2\right)}/q^n$ (representing e.g. the scattering signal of a step target, or the envelope of the integrated signal over a peak of a grating ($w\gg L$), with $\operatorname{erf}(\sigma/\sqrt{2})$-shaped ridge(s)) for monochromatic X-rays and cold plasma $N_{coh,\omega_0}|_{T=0}(q) = m$ (blue dotted); and finite energy bandwidth $\sigma_{\omega}$ and temperature $T_e$ given by $N_{coh}(q) = a\cdot m$ (red dotted); together with the respective measureable signals in the grating peaks above noise $N_{coh}^{eff}(q) = a_{eff}\cdot m$ with $a_{eff}(N_0,N_{incoh})$ from Eqn.~\eqref{eqn:F} and $N_0=N_{coh,\omega_0}|_{T=0}(q)$ (blue solid), or $N_0=\tilde N_0$ (red solid), respectively. 
True values $A_0$, $\sigma_0$ are indicated in the figures, and $n_0=2$ everywhere. \\
First, in \hyperref[subsec:large]{subsection A} below we show the signals for step-like targets or gratings and \emph{large} detector pixel size ($\tilde N_0\propto a$), then in \hyperref[subsec:small]{subsection B} we show the same for gratings and \emph{small} detector pixel size ($\tilde N_0\propto \Gamma^{(0)}$, as discussed in Sec.~\ref{subsubsec:eff}. 
For each data, we add fits using the model function $m$ (dashed and dash-dotted lines with both fixed $n=2$ and free $n$; note: fits start at $q_0=0.01/\mathrm{nm}$), exemplifying the correlation of the temperature $T_e$ and bandwidth $\sigma_{\omega}$ with the smoothness $\sigma$ and geometry $n$ (rectangular rod $n=2$, cylinder $n=3$, sphere $n=4$). 
\vspace{-26pt}
\subsection{Step-function target, or gratings and large detector pixel size $L\,\Delta q \gg 1$: $N_0\propto a$}%
\label{subsec:large}
\subsubsection{Small scattering signal}%
\centerline{\textbf{No background}}
\begin{center}
    \centering
  \makebox[\textwidth][c]{%
  \includegraphics[width=1.3\linewidth]{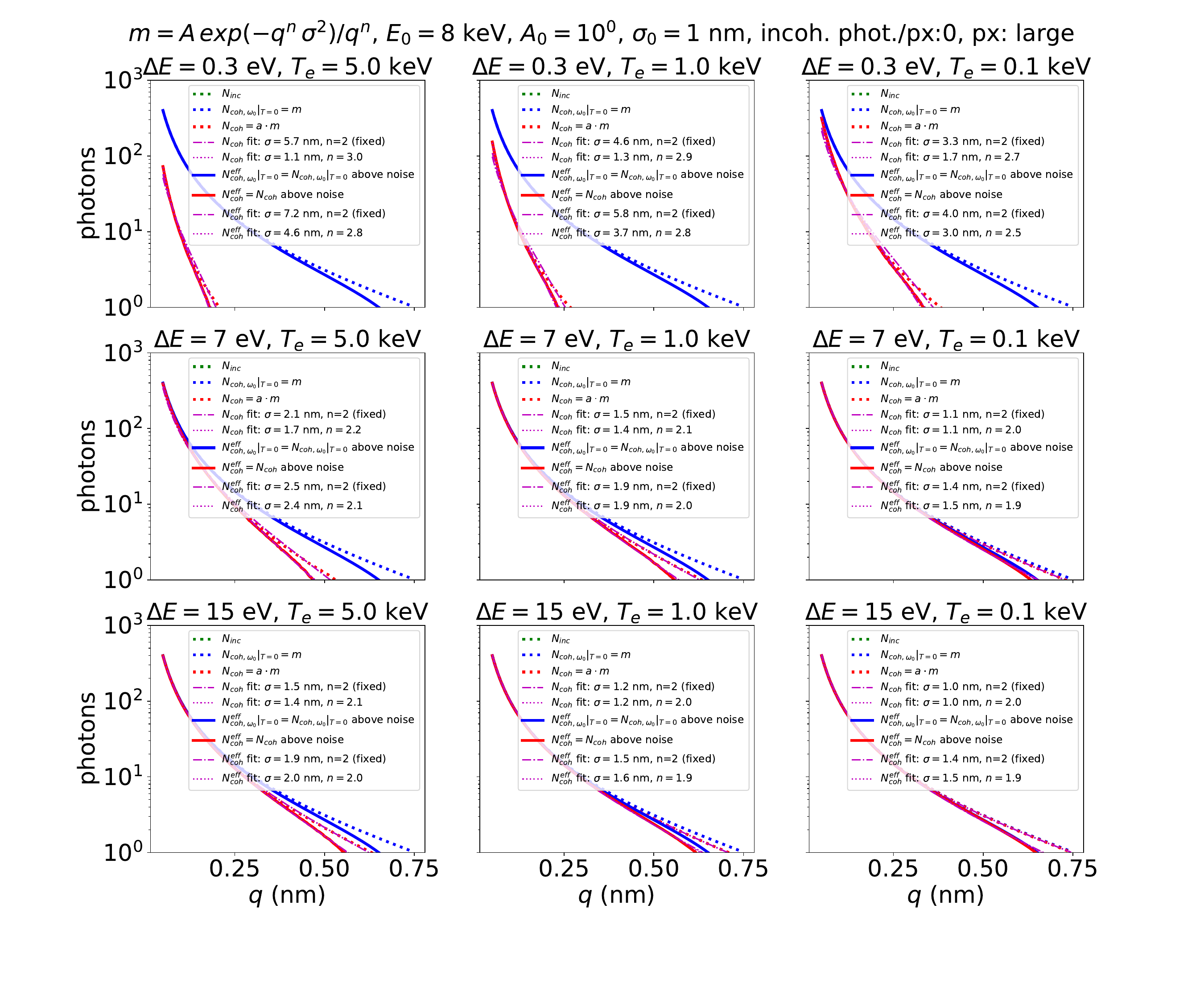}
  }
    \captionof{figure}{Example impact of finite bandwidth on signals and fits for small signal $A_0=1$ and sharp interface $s_0=1\,$nm, ignoring background ($N_{incoh}=0$), and assuming $L\,\Delta q \ll 1$ or single step targets. Dash-dotted and dashed lines are the fit to the respective curves with fixed $n=2$ and free $n$, respectively.
    \label{fig:fitting9}}
\end{center}
\begin{center}
    \centering
  \makebox[\textwidth][c]{%
    \includegraphics[width=1.3\linewidth]{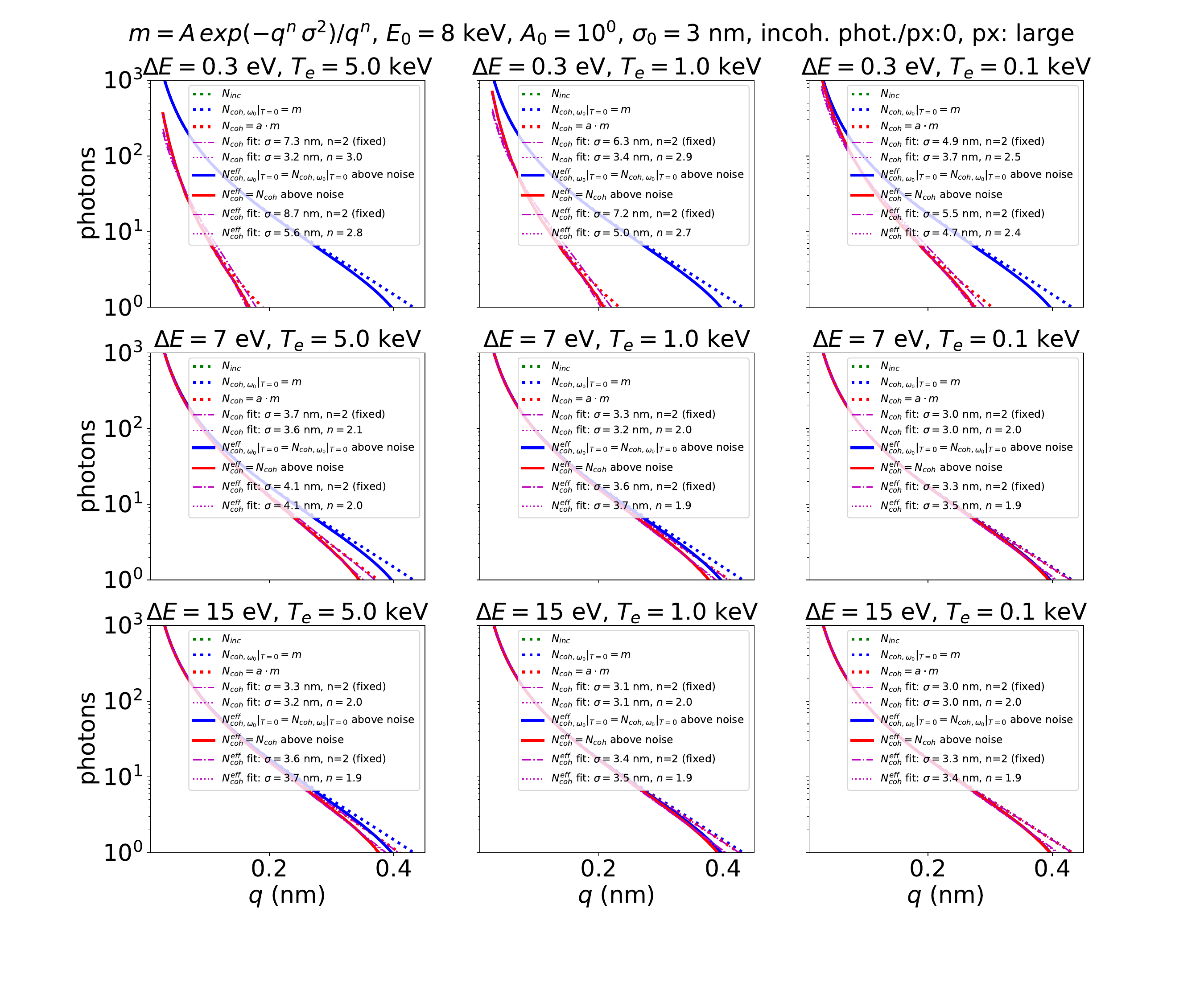}
    }
    \captionof{figure}{Example impact of finite bandwidth on signals and fits for small signal $A_0=1$ and smoother interface $s_0=3\,$nm, ignoring background ($N_{incoh}=0$), and assuming $L\,\Delta q \ll 1$ or single step targets. Dash-dotted and dashed lines are the fit to the respective curves with fixed $n=2$ and free $n$, respectively.
    \label{fig:fitting10}}
\end{center}
\clearpage

\centerline{\textbf{Including background}}
\begin{center}
    \centering
  \makebox[\textwidth][c]{%
    \includegraphics[width=1.3\linewidth]{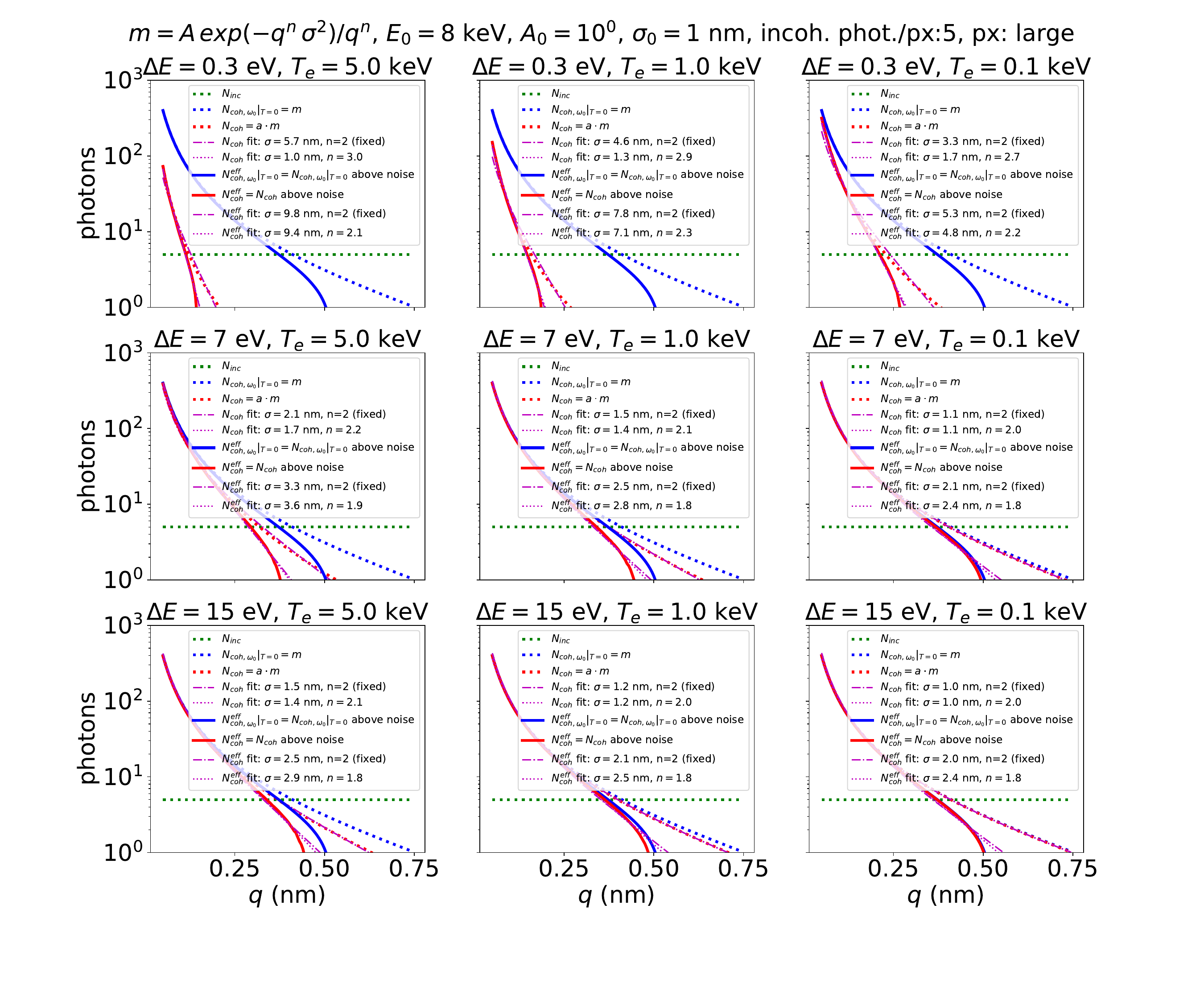}
    }
    \captionof{figure}{Example impact of finite bandwidth on signals and fits for small signal $A_0=1$ and sharp interface $s_0=1\,$nm, assuming 5 counts/px background, and assuming $L\,\Delta q \ll 1$ or single step targets. Dash-dotted and dashed lines are the fit to the respective curves with fixed $n=2$ and free $n$, respectively.
    \label{fig:fitting11}}
\end{center}
\begin{center}
    \centering
  \makebox[\textwidth][c]{%
    \includegraphics[width=1.3\linewidth]{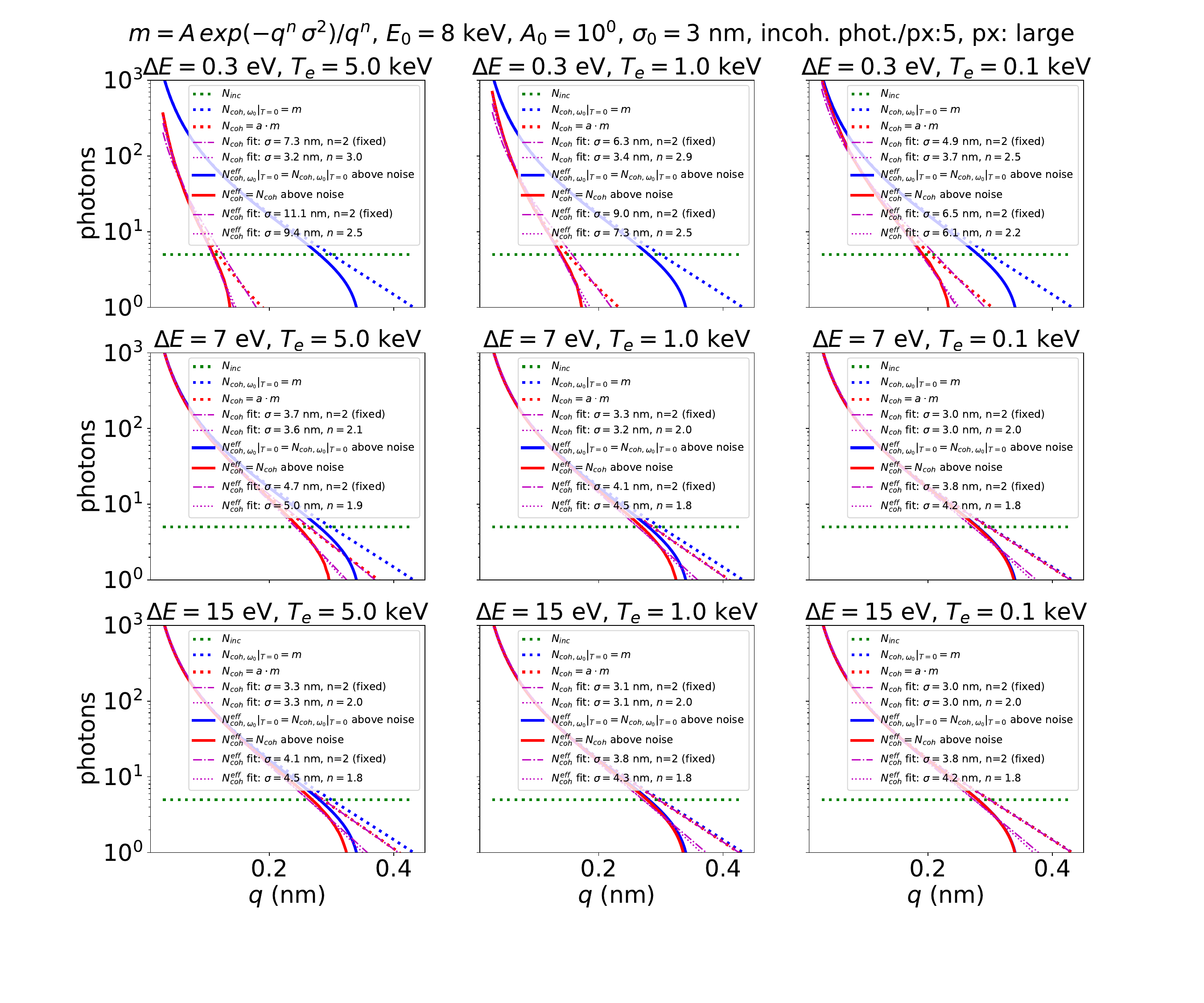}
    }
    \captionof{figure}{Example impact of finite bandwidth on signals and fits for small signal $A_0=1$ and smoother interface $s_0=3\,$nm, assuming 5 counts/px background, and assuming $L\,\Delta q \ll 1$ or single step targets. Dash-dotted and dashed lines are the fit to the respective curves with fixed $n=2$ and free $n$, respectively.
    \label{fig:fitting12}}
\end{center}
\clearpage

\subsubsection{Large scattering signal}
\centerline{\textbf{No background}}
\begin{center}
    \centering
  \makebox[\textwidth][c]{%
    \includegraphics[width=1.3\linewidth]{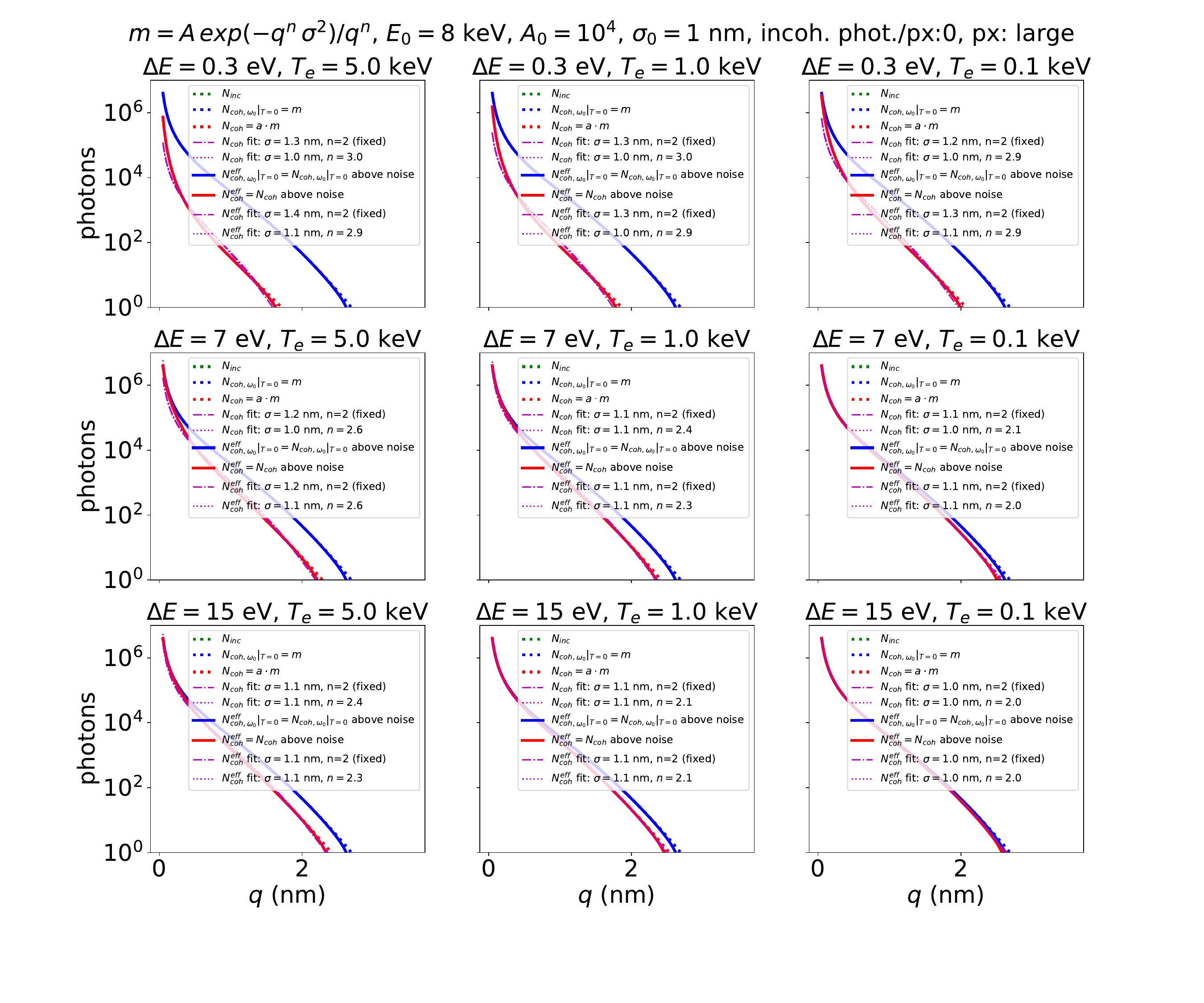}
    }
    \captionof{figure}{Example impact of finite bandwidth on signals and fits for large signal $A_0=1e4$ and sharp interface $s_0=1\,$nm, ignoring background ($N_{incoh}=0$), and assuming $L\,\Delta q \ll 1$ or single step targets. Dash-dotted and dashed lines are the fit to the respective curves with fixed $n=2$ and free $n$, respectively.
    \label{fig:fitting13}}
\end{center}
\begin{center}
    \centering
  \makebox[\textwidth][c]{%
    \includegraphics[width=1.3\linewidth]{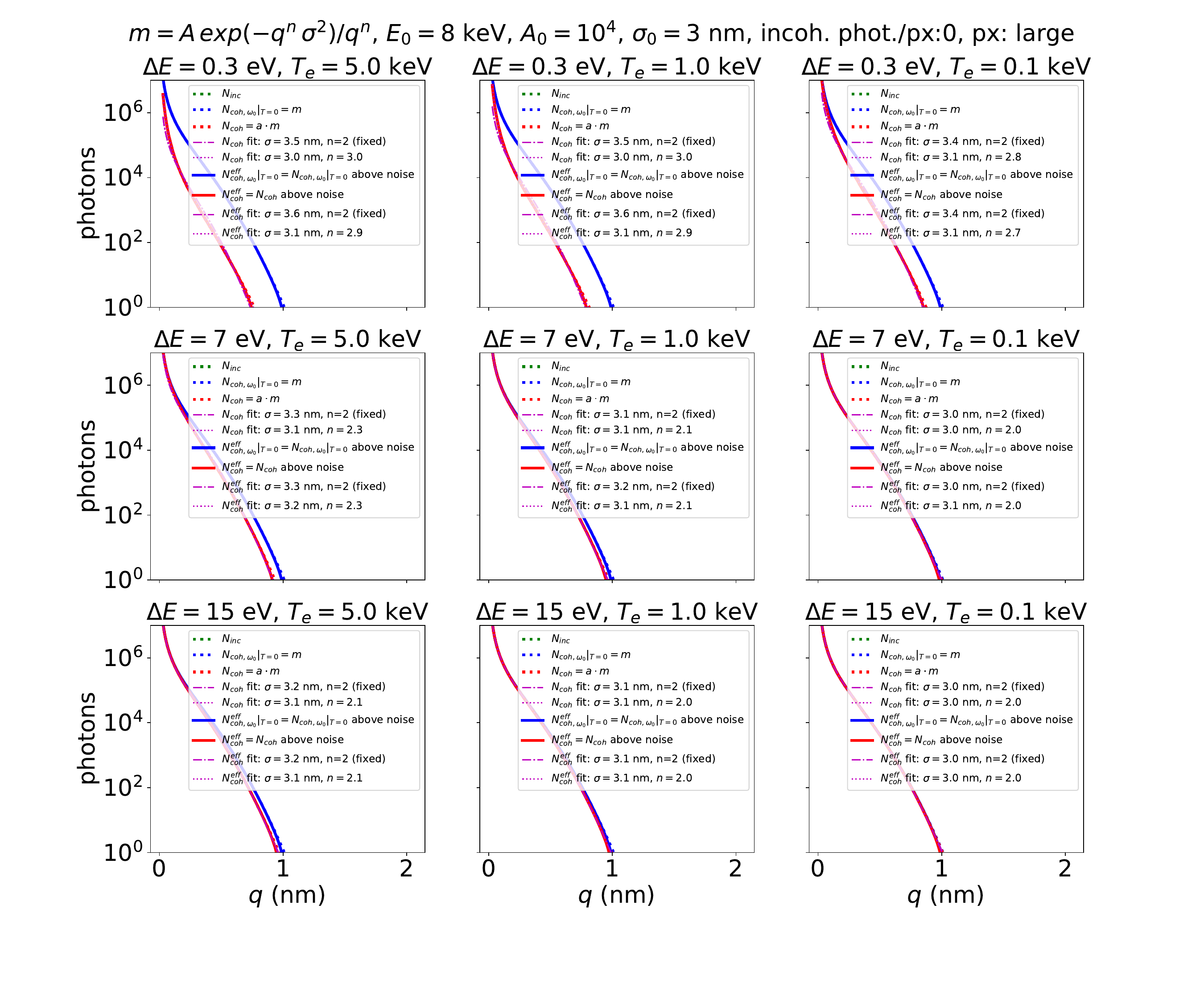}
    }
    \captionof{figure}{Example impact of finite bandwidth on signals and fits for large signal $A_0=1e4$ and smoother interface $s_0=3\,$nm, ignoring background ($N_{incoh}=0$), and assuming $L\,\Delta q \ll 1$ or single step targets. Dash-dotted and dashed lines are the fit to the respective curves with fixed $n=2$ and free $n$, respectively.
    \label{fig:fitting14}}
\end{center}
\clearpage

\centerline{\textbf{Including background}}
\begin{center}
    \centering
  \makebox[\textwidth][c]{%
    \includegraphics[width=1.3\linewidth]{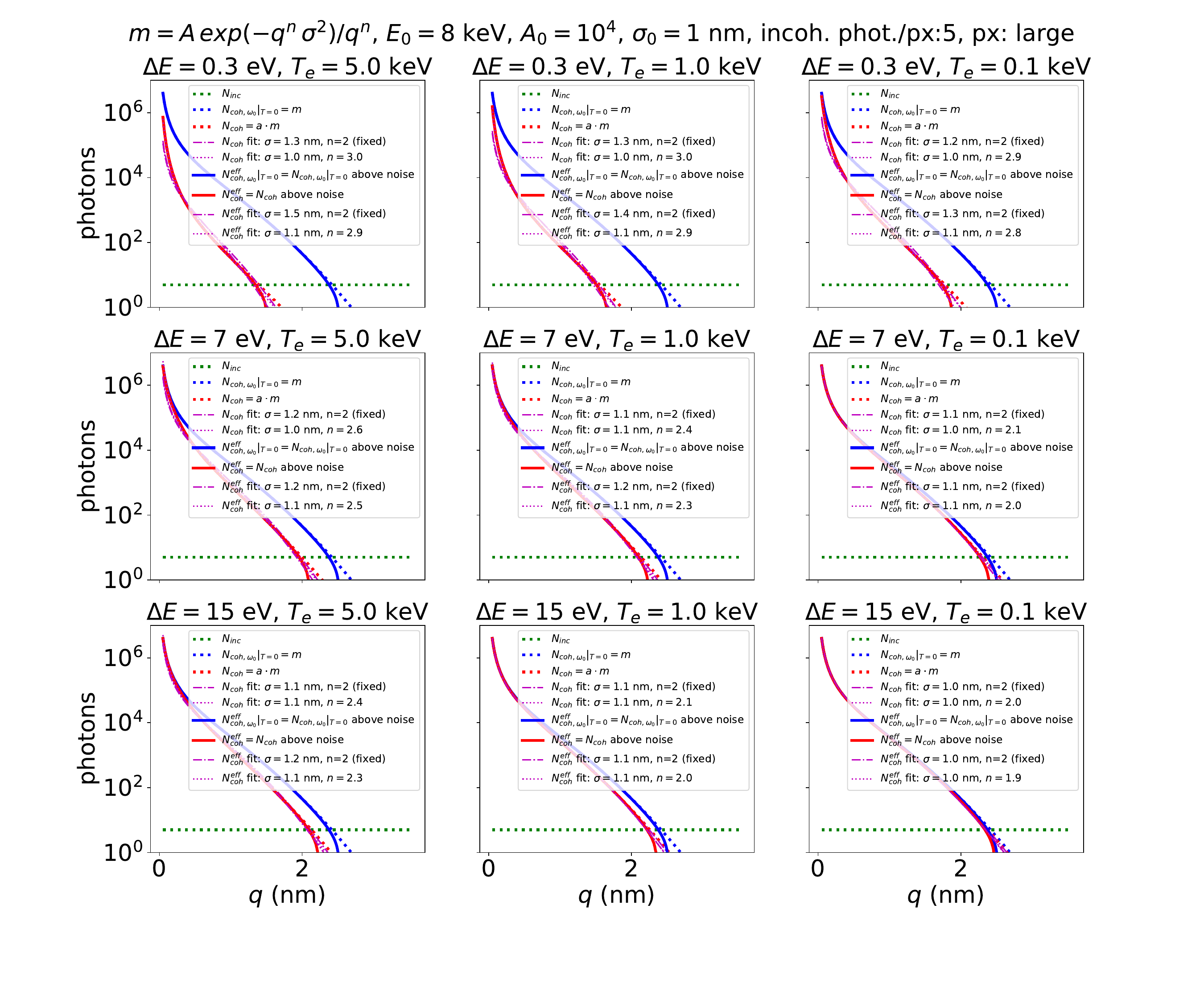}
    }
    \captionof{figure}{Example impact of finite bandwidth on signals and fits for large signal $A_0=1e4$ and sharp interface $s_0=1\,$nm, assuming 5 counts/px background, and assuming $L\,\Delta q \ll 1$ or single step targets. Dash-dotted and dashed lines are the fit to the respective curves with fixed $n=2$ and free $n$, respectively.
    \label{fig:fitting15}}
\end{center}
\begin{center}
    \centering
  \makebox[\textwidth][c]{%
    \includegraphics[width=1.3\linewidth]{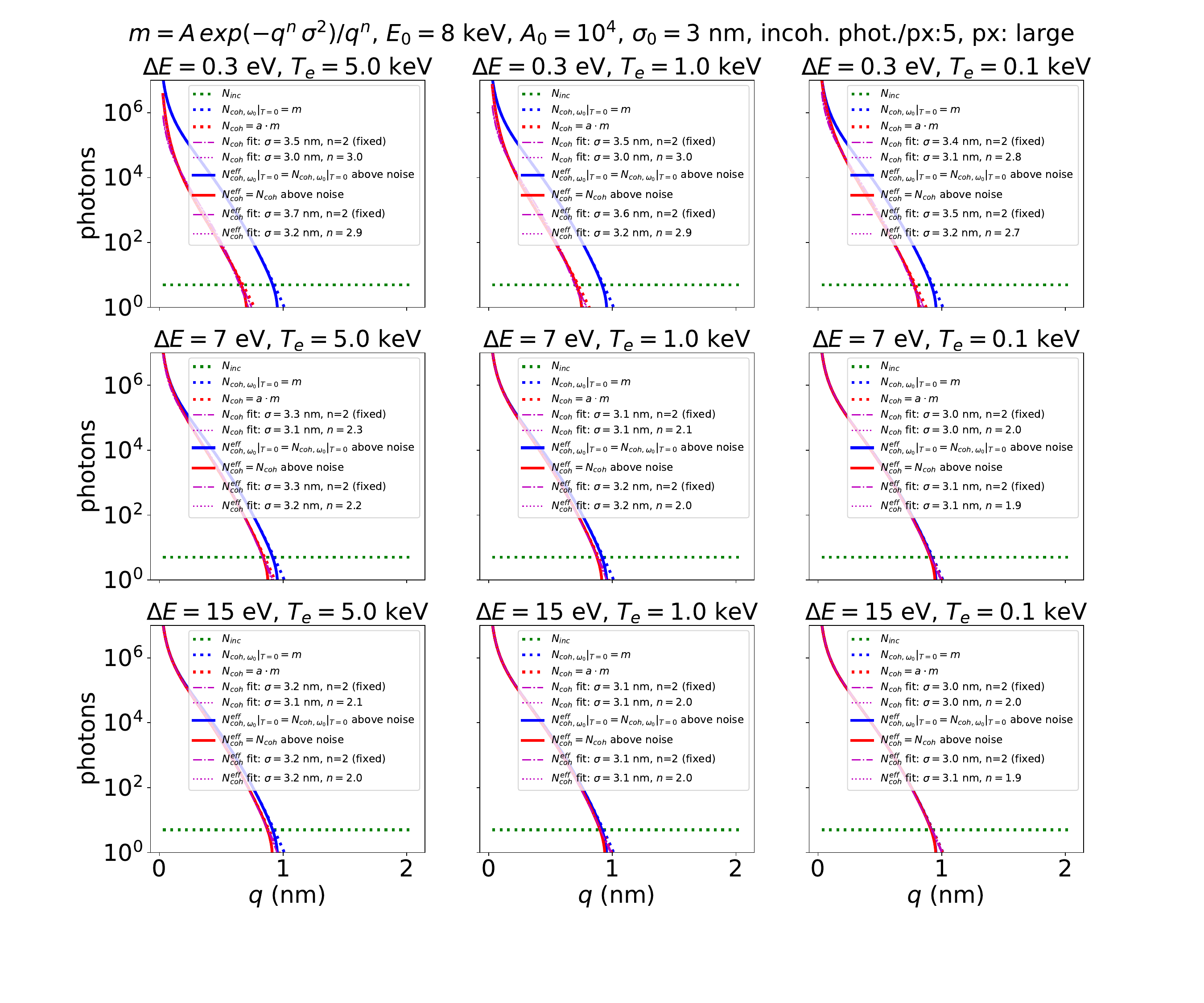}
    }
    \captionof{figure}{Example impact of finite bandwidth on signals and fits for large signal $A_0=1e4$ and smoother interface $s_0=3\,$nm, assuming 5 counts/px background, and assuming $L\,\Delta q \ll 1$ or single step targets. Dash-dotted and dashed lines are the fit to the respective curves with fixed $n=2$ and free $n$, respectively.
    \label{fig:fitting16}}
\end{center}
\clearpage
\subsection{Gratings and small detector pixel size $L\,\Delta q \ll 1$: $N_0\propto\Gamma^{(0)}$}
\label{subsec:small}
\subsubsection{Small scattering signal}
\centerline{\textbf{No background}}
\begin{center}
    \centering
  \makebox[\textwidth][c]{%
    \includegraphics[width=1.3\linewidth]{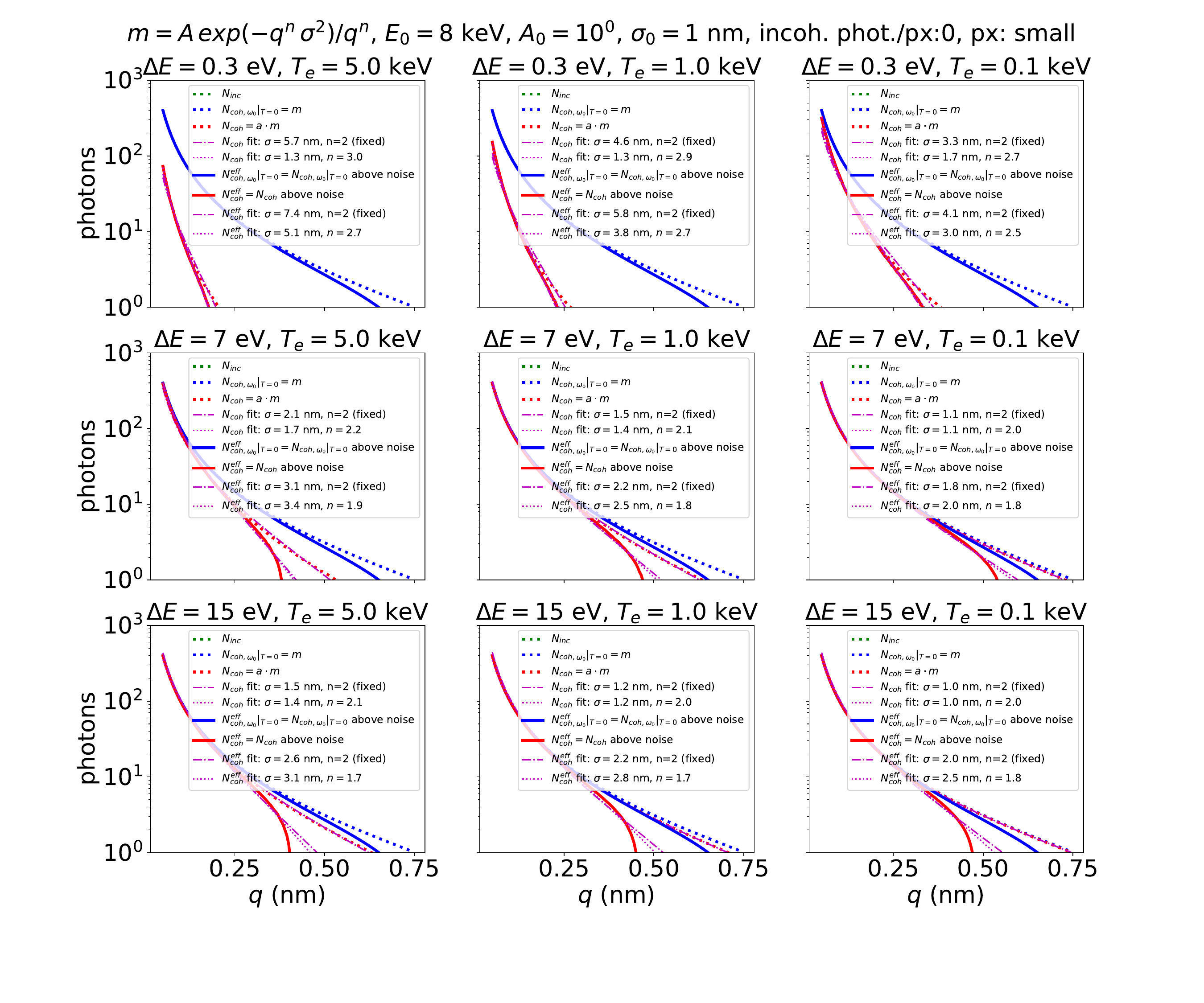}
    }
    \captionof{figure}{Example impact of finite bandwidth on signals and fits for small signal $A_0=1$ and sharp interface $s_0=1\,$nm, ignoring background ($N_{incoh}=0$), and assuming $L\,\Delta q \ll 1$ or single step targets. Dash-dotted and dashed lines are the fit to the respective curves with fixed $n=2$ and free $n$, respectively.
    \label{fig:fitting1}}
\end{center}
\begin{center}
    \centering
  \makebox[\textwidth][c]{%
    \includegraphics[width=1.3\linewidth]{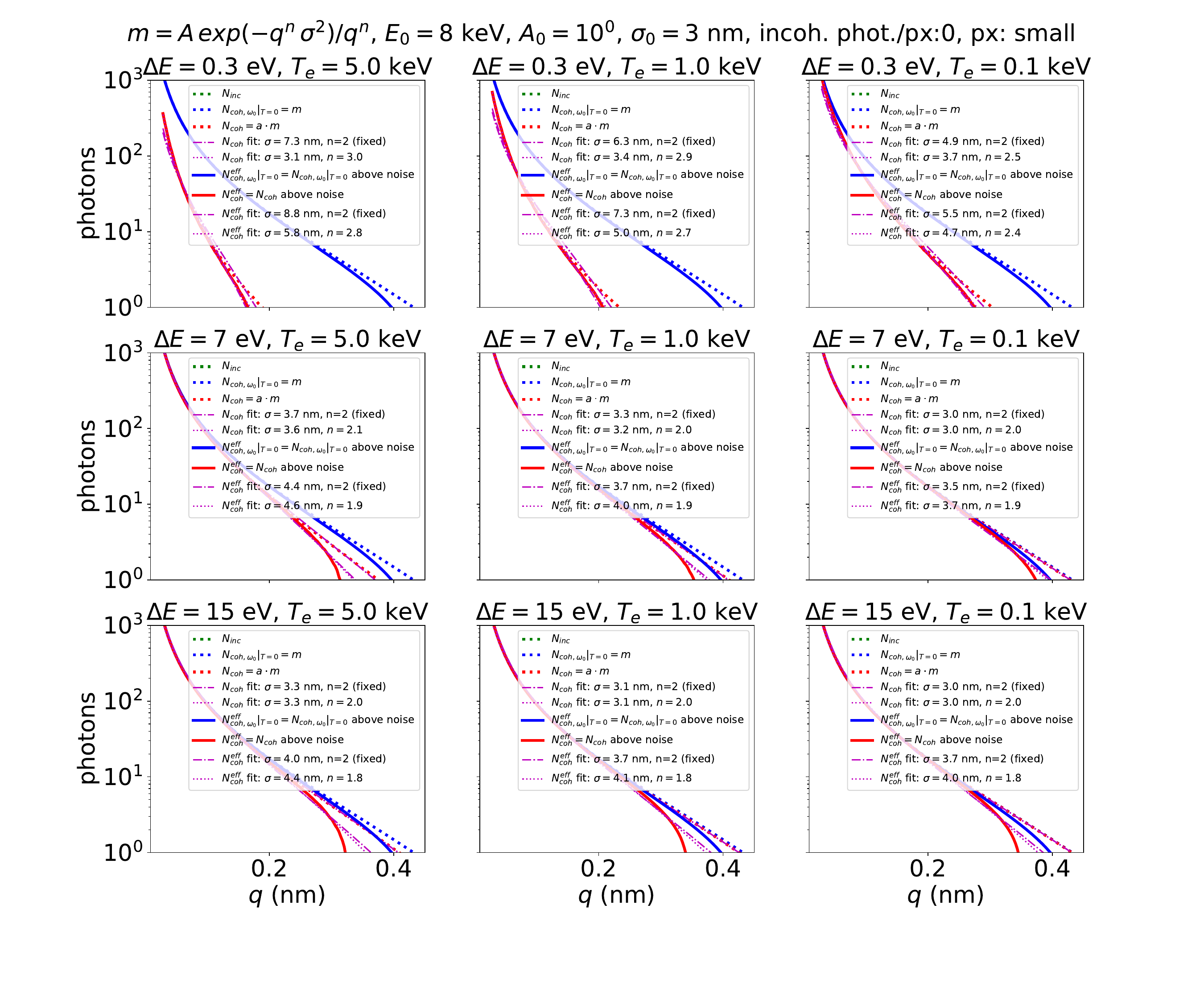}
    }
    \captionof{figure}{Example impact of finite bandwidth on signals and fits for small signal $A_0=1$ and smoother interface $s_0=3\,$nm, ignoring background ($N_{incoh}=0$), and assuming $L\,\Delta q \ll 1$ or single step targets. Dash-dotted and dashed lines are the fit to the respective curves with fixed $n=2$ and free $n$, respectively.
    \label{fig:fitting2}}
\end{center}
\clearpage

\centerline{\textbf{Including background}}
\begin{center}
    \centering
  \makebox[\textwidth][c]{%
    \includegraphics[width=1.3\linewidth]{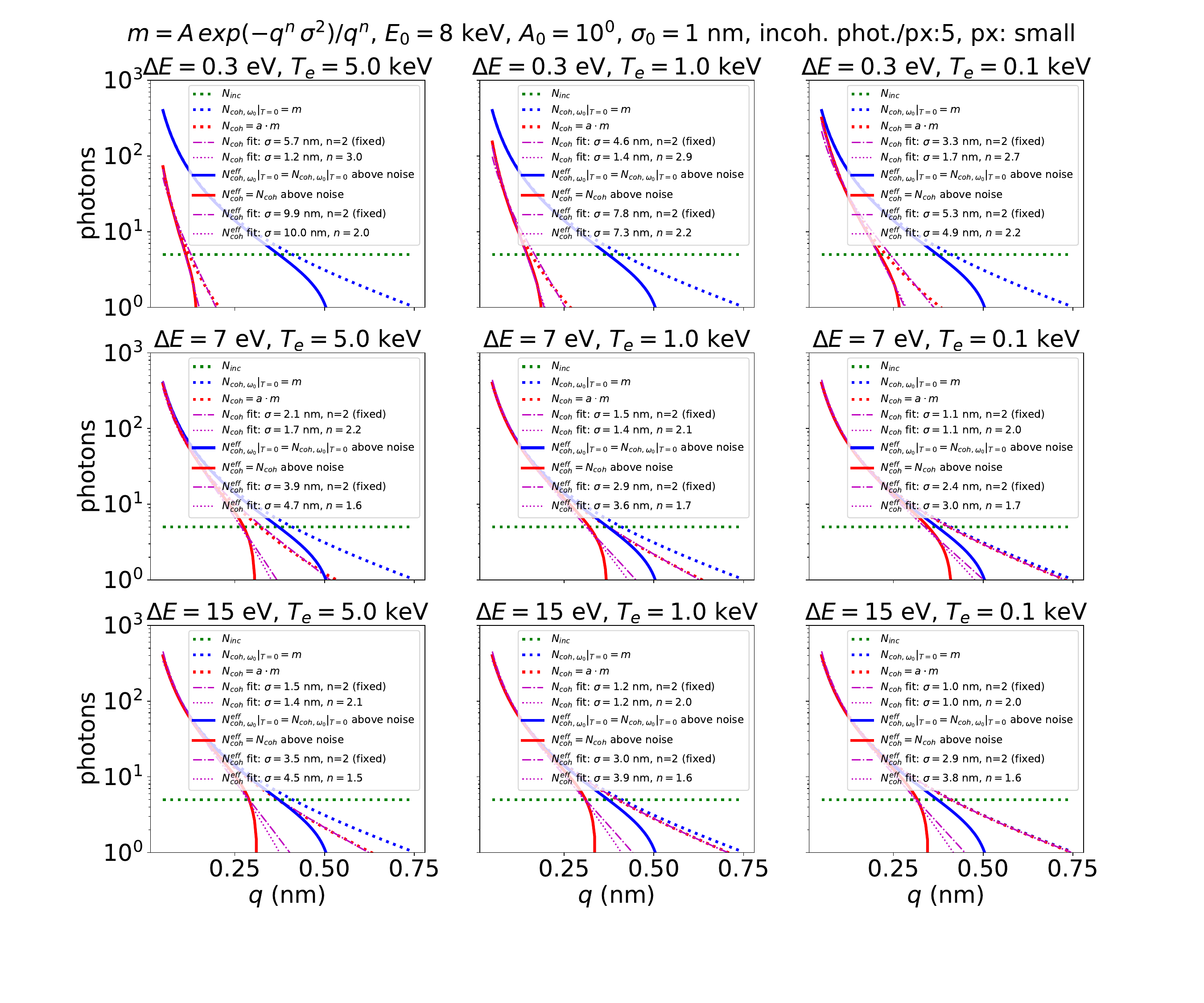}
    }
    \captionof{figure}{Example impact of finite bandwidth on signals and fits for small signal $A_0=1$ and sharp interface $s_0=1\,$nm, assuming 5 counts/px background, and assuming $L\,\Delta q \ll 1$ or single step targets. Dash-dotted and dashed lines are the fit to the respective curves with fixed $n=2$ and free $n$, respectively.
    \label{fig:fitting3}}
\end{center}
\begin{center}
    \centering
  \makebox[\textwidth][c]{%
    \includegraphics[width=1.3\linewidth]{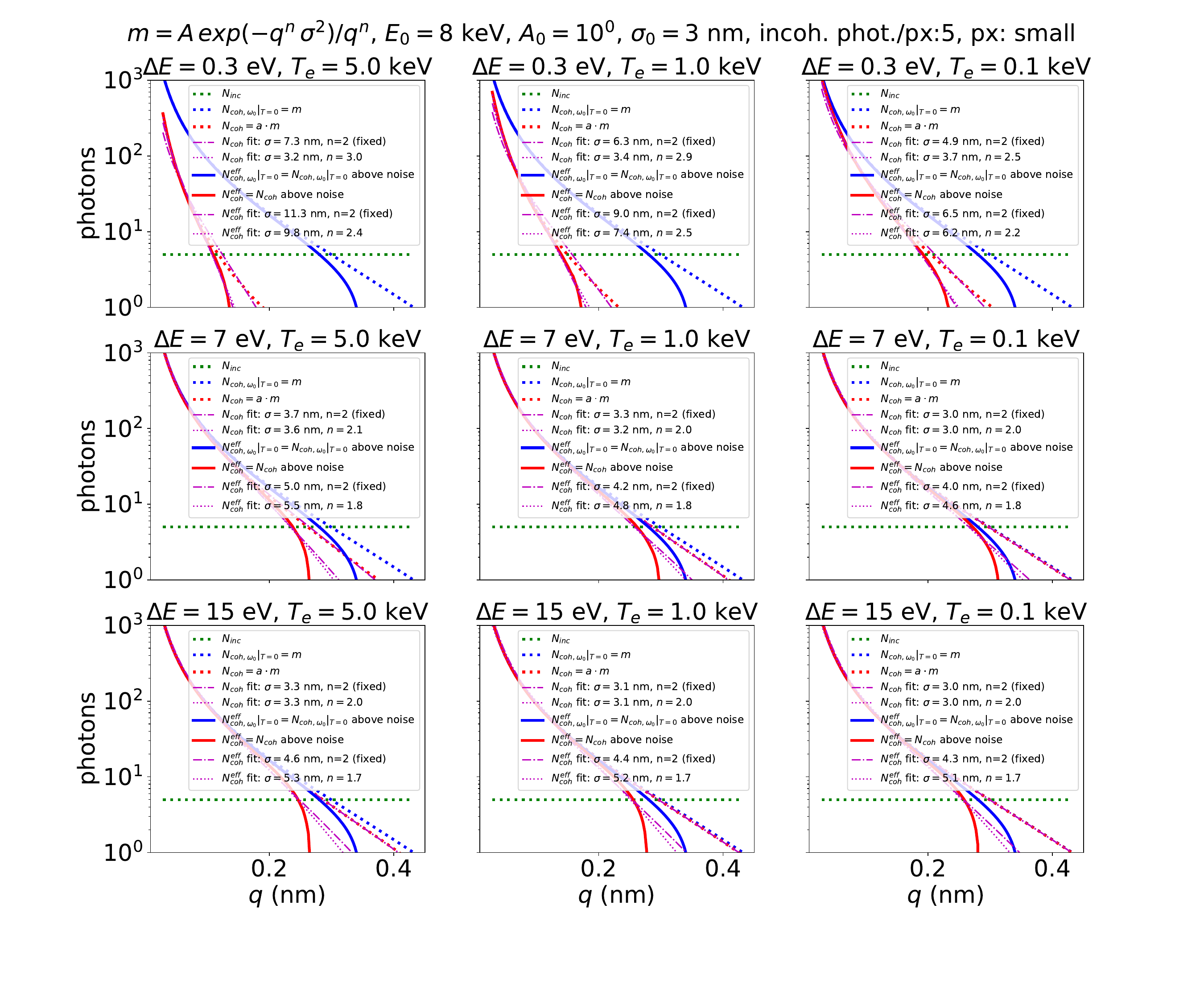}
    }
    \captionof{figure}{Example impact of finite bandwidth on signals and fits for small signal $A_0=1$ and smoother interface $s_0=3\,$nm, assuming 5 counts/px background, and assuming $L\,\Delta q \ll 1$ or single step targets. Dash-dotted and dashed lines are the fit to the respective curves with fixed $n=2$ and free $n$, respectively.
    \label{fig:fitting4}}
\end{center}
\clearpage

\subsubsection{Large scattering signal}
\centerline{\textbf{No background}}
\begin{center}
    \centering
  \makebox[\textwidth][c]{%
    \includegraphics[width=1.3\linewidth]{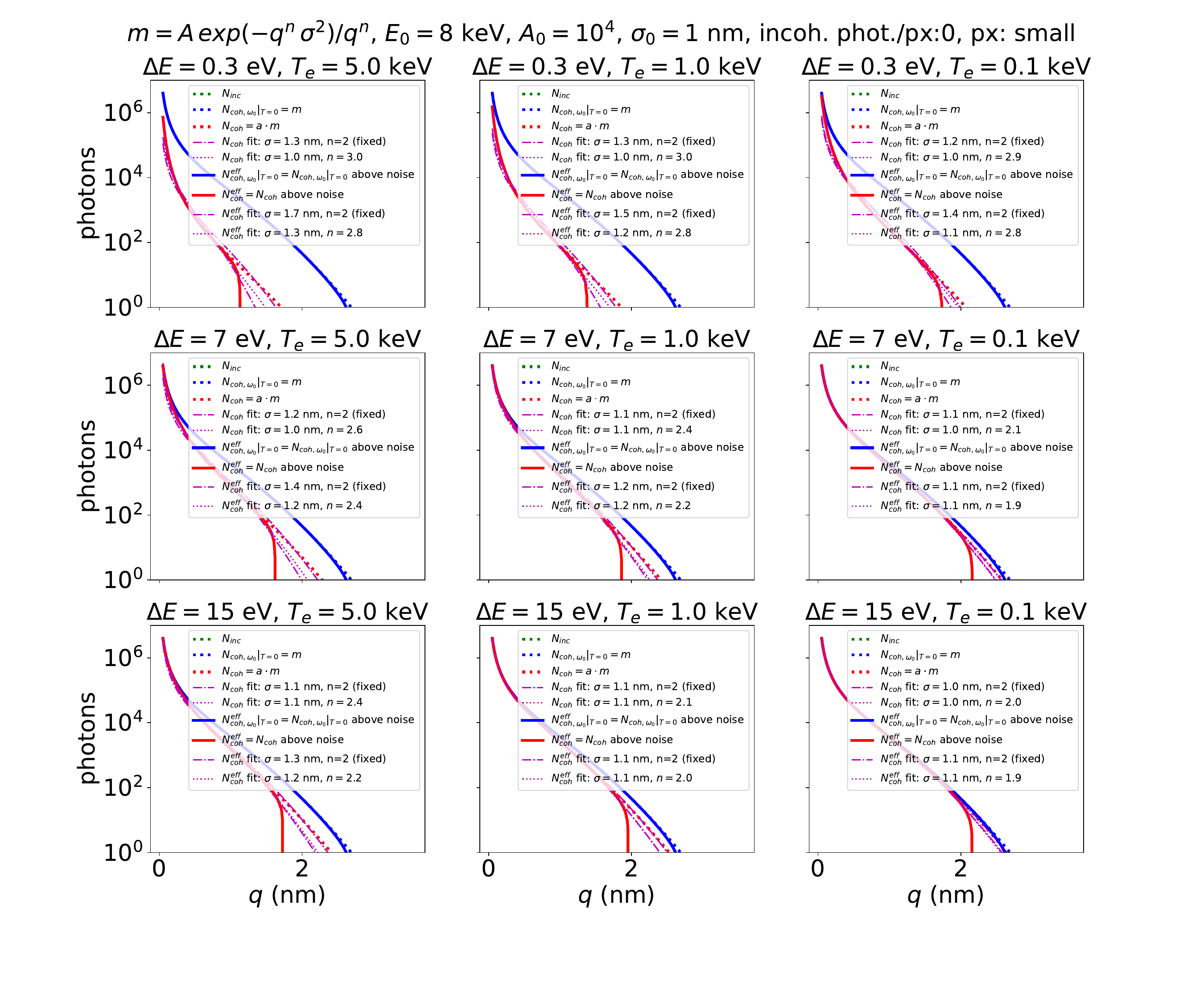}
    }
    \captionof{figure}{Example impact of finite bandwidth on signals and fits for large signal $A_0=1e4$ and sharp interface $s_0=1\,$nm, ignoring background ($N_{incoh}=0$), and assuming $L\,\Delta q \ll 1$ or single step targets. Dash-dotted and dashed lines are the fit to the respective curves with fixed $n=2$ and free $n$, respectively.
    \label{fig:fitting5}}
\end{center}
\begin{center}
    \centering
  \makebox[\textwidth][c]{%
    \includegraphics[width=1.3\linewidth]{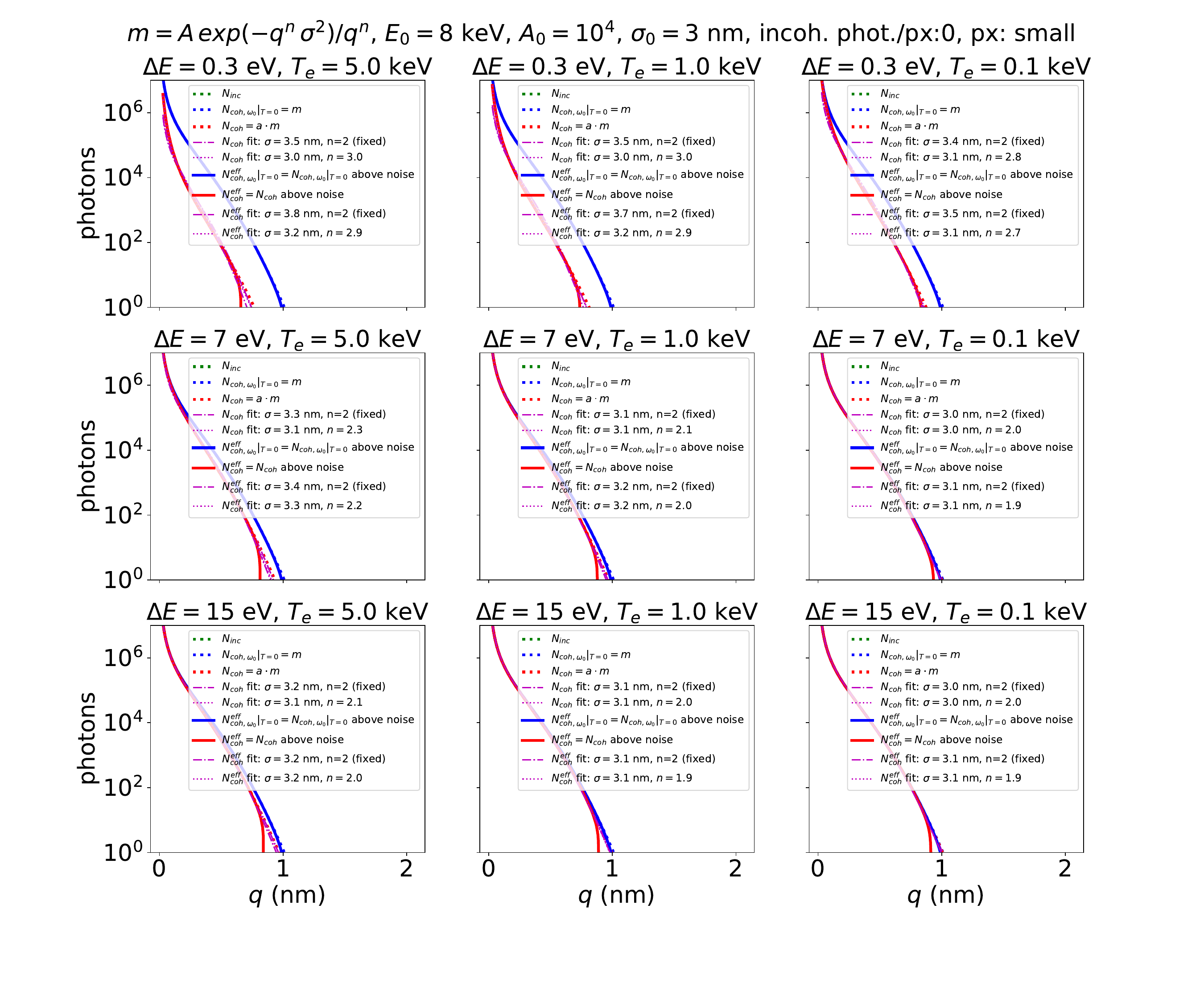}
    }
    \captionof{figure}{Example impact of finite bandwidth on signals and fits for large signal $A_0=1e4$ and smoother interface $s_0=3\,$nm, ignoring background ($N_{incoh}=0$), and assuming $L\,\Delta q \ll 1$ or single step targets. Dash-dotted and dashed lines are the fit to the respective curves with fixed $n=2$ and free $n$, respectively.
    \label{fig:fitting6}}
\end{center}
\clearpage

\centerline{\textbf{Including background}}
\begin{center}
    \centering
  \makebox[\textwidth][c]{%
    \includegraphics[width=1.3\linewidth]{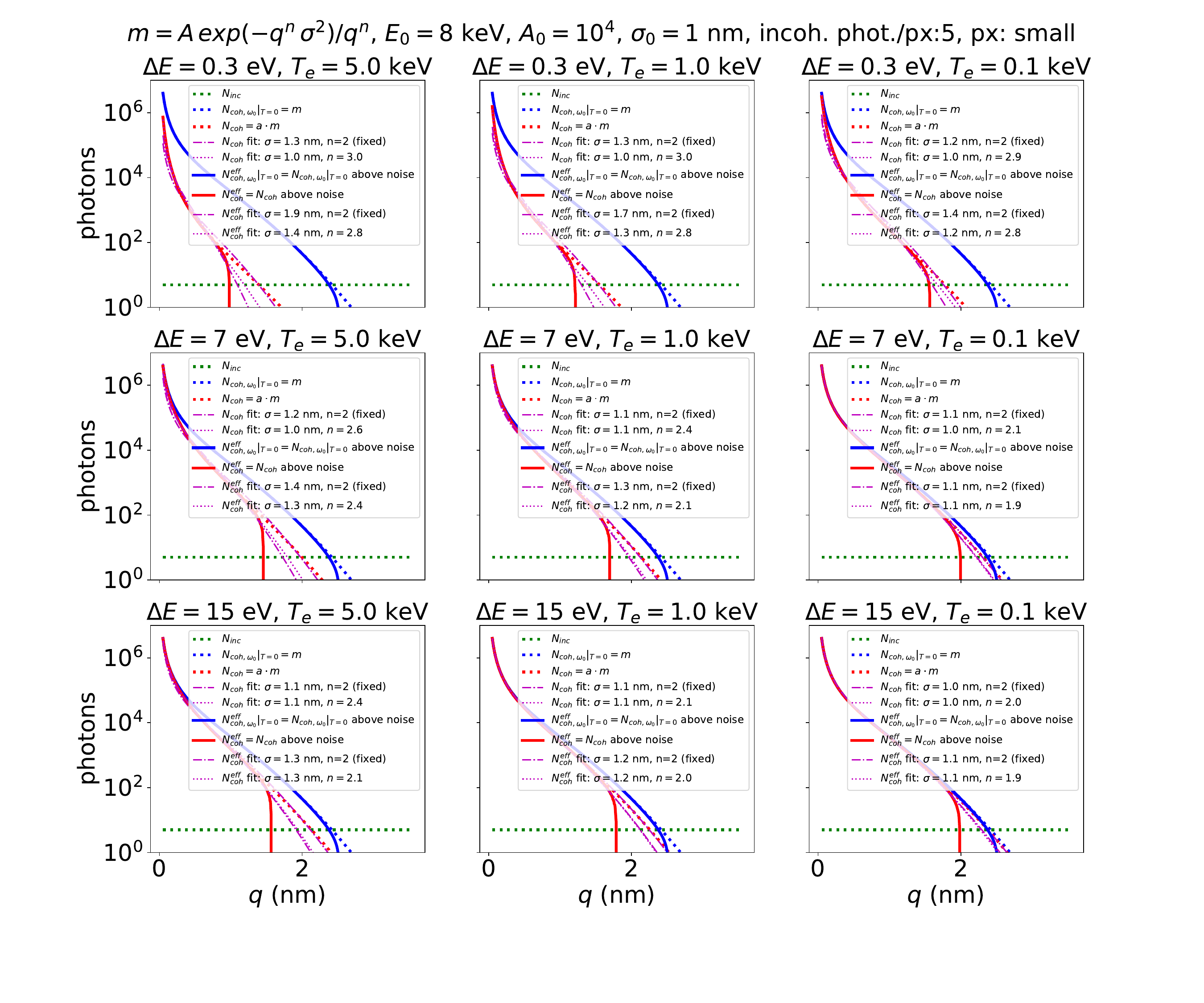}
    }
    \captionof{figure}{Example impact of finite bandwidth on signals and fits for large signal $A_0=1e4$ and sharp interface $s_0=1\,$nm, assuming 5 counts/px background, and assuming $L\,\Delta q \ll 1$ or single step targets. Dash-dotted and dashed lines are the fit to the respective curves with fixed $n=2$ and free $n$, respectively.
    \label{fig:fitting7}}
\end{center}
\begin{center}
    \centering
  \makebox[\textwidth][c]{%
    \includegraphics[width=1.3\linewidth]{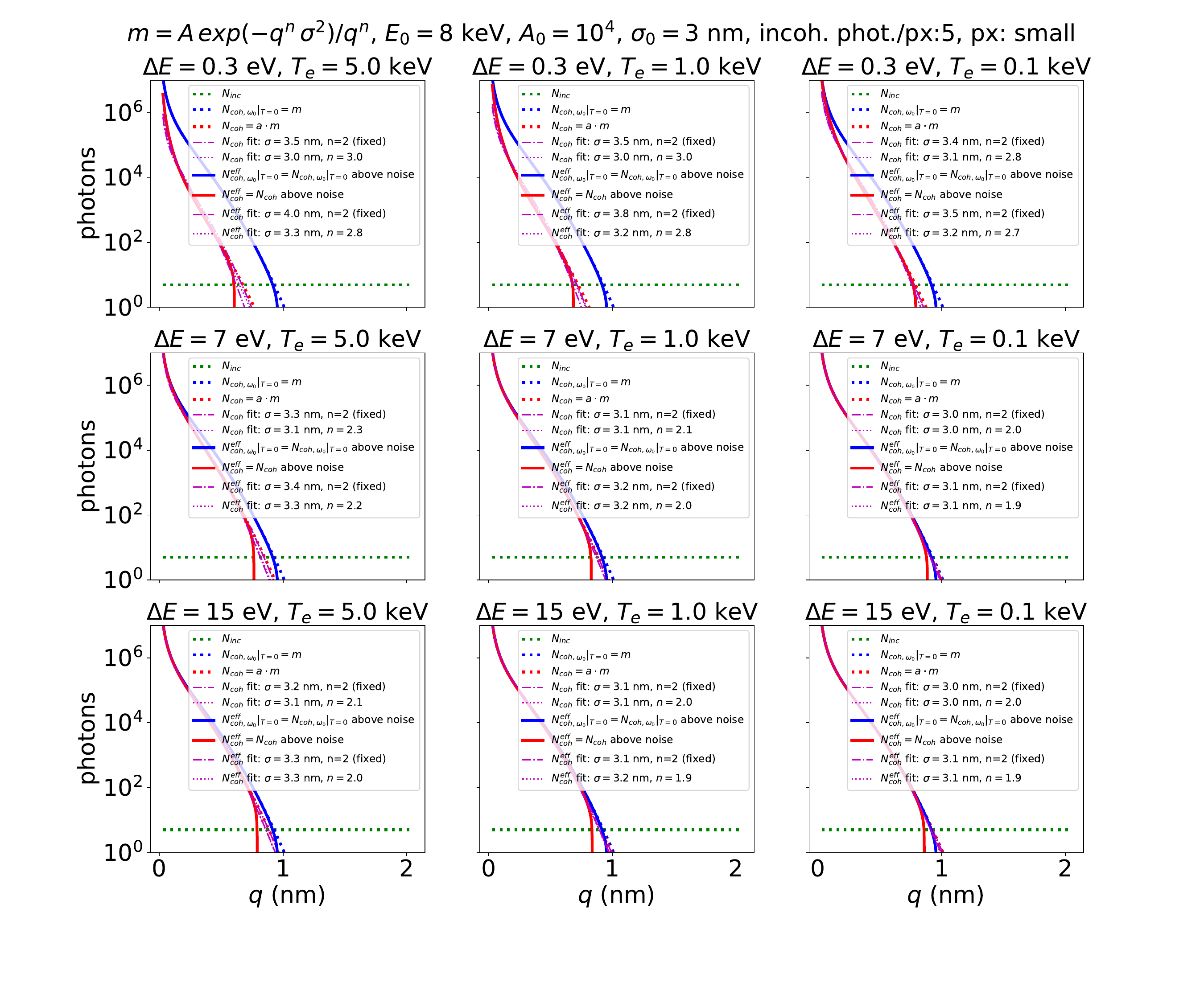}
    }
    \captionof{figure}{Example impact of finite bandwidth on signals and fits for large signal $A_0=1e4$ and smoother interface $s_0=3\,$nm, assuming 5 counts/px background, and assuming $L\,\Delta q \ll 1$ or single step targets. Dash-dotted and dashed lines are the fit to the respective curves with fixed $n=2$ and free $n$, respectively.
    \label{fig:fitting8}}
\end{center}
\clearpage
\section*{Appendix D}


\begin{center}
    \centering
  \makebox[\textwidth][c]{%
    \includegraphics[width=1\linewidth]{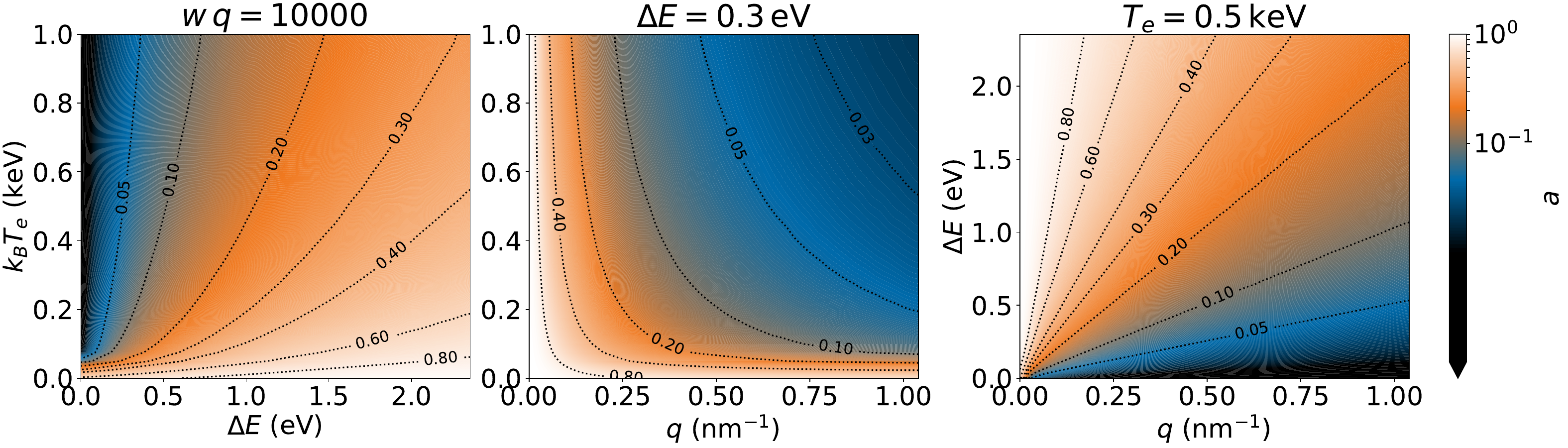}
    }
    \captionof{figure}{Reduction of fluence $\int_{-\infty}^{+\infty}
    I_l(k)dk/\int_{-\infty}^{+\infty}
    I_{l,\omega_0}|_{T=0}dk=a$ (see Eqn.~\eqref{eqn:normalization} with $a$ from~\eqref{eqn:a}) for parameters relevant for seeded beams.  
    \label{fig:wires_small}}
\end{center}

\begin{center}
    \centering
  \makebox[\textwidth][c]{%
    \includegraphics[width=\linewidth]{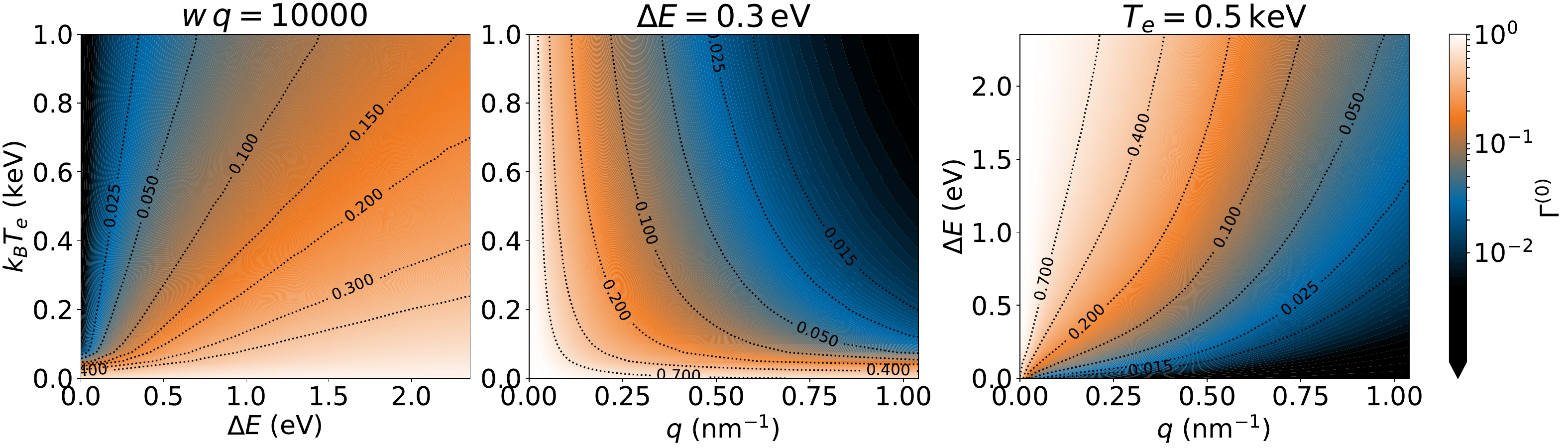}
    }
    \captionof{figure}{Effect of finite coherence and Doppler shift on grating peak height $\Gamma(0)$ (Eqn.~\eqref{eqn:grating_amplitude} with $a,L$ from Eqns.~\eqref{eqn:a},~\eqref{eqn:L}) for parameters relevant for seeded beams. 
    \label{fig:gratings_small}}
\end{center}
\end{widetext}

\end{document}